\documentclass[%
  twocolumn,
  aps,
  prb,
  amsmath,
  amssymb,
  superscriptaddress,
  nofootinbib,
  floatfix
]{revtex4-1}

\usepackage{graphicx}
\usepackage{hyperref}
\usepackage{dcolumn}
\usepackage{bm}
\usepackage{xcolor}

\newcommand{\s}{\sum\limits}

\newcommand{\be}{\begin{equation}}
\newcommand{\e}{\end{equation}}
\newcommand{\beml}{\begin{subequations}}
\newcommand{\eml}{\end{subequations}}
\newcommand{\beq}{\begin{eqnarray}}
\newcommand{\eq}{\end{eqnarray}}
\newcommand{\ba}{\begin{array}}
\newcommand{\ea}{\end{array}}
\newcommand{\bpm}{\begin{pmatrix}}
\newcommand{\epm}{\end{pmatrix}}
\newcommand{\bc}{\begin{cases}}
\newcommand{\ec}{\end{cases}}
\newcommand{\lt}{\left}
\newcommand{\rt}{\right}
\newcommand{\n}{\nonumber}
\newcommand{\la}{\langle}
\newcommand{\ra}{\rangle}

\newcommand{\bb}{\boldsymbol}
\newcommand{\h}{^\dagger}
\newcommand{\0}{^{\phantom{\dagger}}}

\begin{document}

\title{Spin-orbit torques in a Rashba honeycomb antiferromagnet}

\author{Robert Sokolewicz}
\affiliation{Institute for Molecules and Materials, Radboud University Nijmegen, NL-6525 AJ Nijmegen, the Netherlands}

\author{Sumit Ghosh}
\affiliation{Physical Science and Engineering Division, King Abdullah University of Science and Technology, Thuwal 23955, Saudi Arabia}
\affiliation{Peter Gr\"unberg Institut, Forschungszentrum J\"ulich, 52425 J\"ulich, Germany}

\author{Dmitry Yudin}
\affiliation{Skolkovo Institute of Science and Technology, Moscow 121205, Russia}
\affiliation{Department of Physics and Astronomy, Uppsala University, Box 516, SE-751 20, Uppsala, Sweden}

\author{Aur\'elien Manchon}
\affiliation{Physical Science and Engineering Division, King Abdullah University of Science and Technology, Thuwal 23955, Saudi Arabia}

\author{Mikhail Titov}
\affiliation{Institute for Molecules and Materials, Radboud University Nijmegen, NL-6525 AJ Nijmegen, the Netherlands}
\affiliation{ITMO University, Saint Petersburg 197101, Russia}

\date{\today}

\begin{abstract}
Recent experiments on switching antiferromagnetic domains by electric current pulses have attracted a lot of attention to spin-orbit torques in antiferromagnets. In this work, we employ the tight-binding model solver, kwant,  to compute spin-orbit torques in a two-dimensional antiferromagnet on a honeycomb lattice with strong spin-orbit interaction of Rashba type. Our model combines spin-orbit interaction, local $s$-$d$-like exchange, and scattering of conduction electrons on on-site disorder potential to provide a microscopic mechanism for angular momentum relaxation. We consider two versions of the model: one with preserved and one with broken sublattice symmetry. A non-equilibrium staggered polarization, that is responsible for the so-called N\'eel spin-orbit torque, is shown to vanish identically in the symmetric model but may become finite if sublattice symmetry is broken. Similarly, anti-damping spin-orbit torques vanish in the symmetric model but become finite and anisotropic in a model with broken sublattice symmetry. As expected, anti-damping torques also reveal a sizable dependence on impurity concentration. Our numerical analysis also confirms symmetry classification of spin-orbit torques and strong torque anisotropy due to in-plane confinement of electron momenta. 
\end{abstract}

\maketitle

\section{Introduction}

Ferromagnetic spintronics has greatly contributed to microelectronic technology in the last few decades \cite{Bader2010, Sinova2012, Bhatti2017}. One of the practical results was the development of magnetic memories with purely electronic write-in and read-out processes as an alternative to existing solid state drive technologies \cite{Kent2015, Sato2018}. An increasing demand for ever higher performance computation and ever faster big data analytics has sparked recently the interest to antiferromagnetic spintronics \cite{MacDonald2011, Gomonay2014, Wadley2016, Jungwirth2016AFMreview, Baltz2018, Jungwirth2018, Hoffman2018}, i.\,e. to the usage of much more subtle antiferromagnetic order parameter to store and process information. This idea is driven primarily by the expectation that antiferromagnetic materials may naturally allow for up to THz operation frequencies \cite{Gomonay2016AFM, Olejnik2018, Jungwirth2018} in sharp contrast to ferromagnets whose current-induced magnetization dynamics is fundamentally limited to GHz frequency range. 
  
The best efficiency in electric switching of magnetic domains is achieved in systems involving materials with at least partial spin-momentum locking \cite{Fina2016} due to sufficiently strong spin-orbit interaction. The latter is responsible for the so-called spin-orbit torques on magnetization that are caused by sizable non-equilibrium spin polarization induced by an electron flow \cite{Brataas2012, Hals2013, Zelezny2014, 2014MokrousovSOT, Ghosh2017, SmejkalAFM_2017, Zelezny2018, Zhou2018, Manchon2018, Moriyama2018, Li2019, Chen2019, Zhou2019, Zhou2019a, Bodnar2018}.  

Recently, spin-orbit-torque-driven electric switching of the N\'eel vector orientation has been predicted \cite{Zelezny2014} and discovered in non-centrosymmetric crystals such as CuMnAs \cite{Wadley2016, Fina2016, Zelezny2018, Saidl2017} and Mn$_2$Au \cite{Barthem2013, Jordan2015, Bhattacharjee2018}. Even though many antiferromagnetic compounds are electric insulators \cite{Pandey2017}, which limits the range of their potential applications, e.\,g., for spin injection \cite{Tshitoyan2015}, the materials like CuMnAs and Mn$_2$Au possess  semi-metal and metal properties, inheriting strong spin-orbit coupling and sufficiently high conductivity. These materials also give rise to collective mode excitations in THz range \cite{Bhattacharjee2018}. 

Spin-orbit torque in antiferromagnets have been investigated theoretically using Kubo-Streda formula in the case of two-dimensional (2D) Rashba gas, as well as in tight-binding models of Mn$_2$Au \cite{Zelezny2014, Zelezny2017}. These pioneering works describe the spin-orbit torques based on their influence on the current-driven dynamics as predicted by Landau-Lifshitz-Gilbert equation. In particular, Zelezny {\it et al.} proposed that only a staggered field-like torque and an non-staggered anti-damping torque can trigger current-driven antiferromagnetic THz switching and excitation in antiferromagnets (see e.\,g., \onlinecite{Faerbu1995, Cheng2016, Khymyn2017}). This analysis served as a basis to further investigation of spin-orbit torques in heterostructures \cite{ManchonJPCM2017, Ghosh2017, Ghosh2019}. Furthermore, a symmetry group analysis has been developed to predict the form of these two components based on the magnetic point groups of the antiferromagnets \cite{Zelezny2017, Watanabe2018}. 
 
In the context of current-driven antiferromagnetic domain wall \cite{Hals2011} and skyrmion motion \cite{Barker2016, Zhang2016}, the N\'eel vector dynamics has been modeled within the phenomenological treatment of the Landau-Lifshitz-Gilbert equation pioneered by Slonczewski \cite{SLONCZEWSKI1996}. In this phenomenological approach the torques on magnetization derived above have been simply postulated \cite{Gomonay2016AFM, Shiino2016, Akosa2018, Tomasello2017}. 

While most of the above studies have focused on the existence and influence of the staggered field-like and non-staggered anti-damping torques, in the present work, we do compute all types of spin-orbit torques from a long-scale numerical analysis in a diffusive transport regime. Our study of spin-orbit torques is based on the numerical computation of non-equilibrium spin polarizations for an effective $s$-$d$ type model in a two-dimensional (2D) honeycomb antiferromagnet with Rashba spin-orbit coupling and on-site disorder potential.  Our results stress the importance of anisotropy of both field-like and anti-damping like spin-orbit torques due to 2D confinement of conduction electrons. We also find, surprisingly, that both the staggered field-like and non-staggered anti-damping torques vanish identically in a model with an $s$-$d$ exchange coupling that is the same on two antiferromagnetic sublattices. In contrast, the model with strongly asymmetric $s$-$d$ exchange couplings leads to finite anti-damping torques while the entire notion of staggered torques becomes largely irrelevant in such an asymmetric  model. 

\begin{figure}
\centering
\centerline{\includegraphics[width=\columnwidth]{{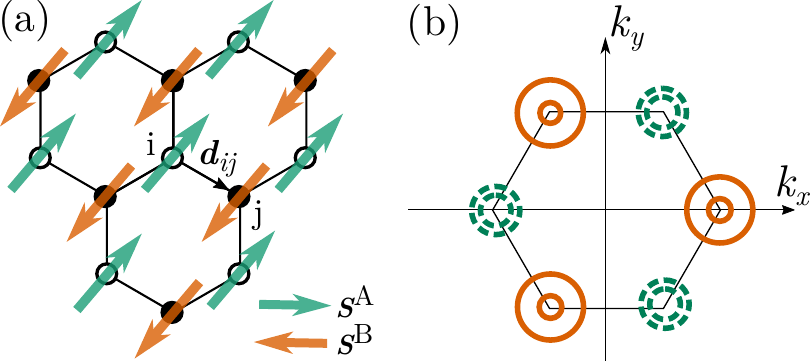}}}
\caption{
Left panel (a): a two-dimensional honeycomb lattice hosting anti-parallel localized magnetic moments $\bb{S}^\textrm{A}$ and $\bb{S}^\textrm{B}$ that induce opposite exchange potentials on A and B sublattice. The vector $\bb{d}_{ij}$ is directed from an A-site $i$ to one of the three nearest-neighbor B-sites $j$. Right panel (b): Fermi surfaces (strongly enlarged) in quasi-momentum space for the Fermi energy $E=0.3\,w$, couplings $\lambda = 0.05\,w$, $\Delta=JS=0.1\,w$, and $\theta=\pi/2$. Solid (orange) and dashed (green) lines indicate $K$ and $K'$ valleys. 
}
\label{fig:lattice}
\end{figure}

\section{Model}

In this paper, we develop a numerical framework for the microscopic analysis of spin-orbit torques that is realized with the help of the tight-binding model solver, kwant \cite{Groth2014}. Our methodology is illustrated using $s$-$d$ type model for a 2D antiferromagnet on the honeycomb lattice depicted in Fig.~\ref{fig:lattice}a.  
To be more specific, we consider the Kane-Mele tight-binding model for conduction electrons \cite{Kane2005, Qiao2012} with the Hamiltonian
\be
\label{ham}
H_0=H_\mathrm{tb}+H_\textrm{R}+H_\mathrm{sd},
\e
where the first term describes the nearest-neighbor hopping, 
\be
\label{hop}
H_{\textrm{tb}}=-w \s_{\la i, j \ra} \s_{\sigma} c\h_{i\sigma} c\0_{j\sigma},
\e
with the hopping energy $w$. The operators $c\h_{i\sigma}$ ($c\0_{i\sigma}$) are the standard creation (annihilation) operators for a fermion on the lattice site $i$ with the spin index $\sigma$. 

The term $H_\textrm{R}$ describes spin-orbit interaction of Rashba type that is responsible for spin-orbit torque on magnetization. This spin-orbit interaction is represented by the term
\be
\label{rashba}
H_\textrm{R}=\frac{i\lambda}{a} \s_{\la i,j\ra}\s_{\sigma\sigma'}
\hat{\bb{z}}\cdot(\bm{\sigma}\times{\bb{d}}_{ij})_{\sigma\sigma'}c_{i\sigma}^\dagger c\0_{j\sigma'},
\e
with the unit vector $\hat{\bb{z}}$ directed perpendicular to the 2D crystal plane. The notation $\bb{\sigma}=(\sigma_x,\sigma_y,\sigma_z)$ represents the three-dimensional vector of Pauli matrices, the vectors $\bb{d}_{ij}$ connect neighboring sites as shown in Fig.~\ref{fig:lattice}a, and $\lambda$ is the Rashba-spin-orbit strength. For any site $i$ on the sublattice $A$ we have three such vectors:
\be
\bb{d}_{1}= a \bpm 0 \\ 1 \epm, \quad \bb{d}_{2}= \frac{a}{2} \bpm \sqrt{3} \\ -1 \epm , \quad \bb{d}_{3}= -\frac{a}{2} \bpm \sqrt{3}  \\ 1 \epm,
\e
where $a$ is the length of the bond between $A$ and $B$.

Finally, the $s$-$d$-like exchange interaction between localized magnetic moments $\bb{S}_i$ and spins of conduction electrons is described by the term
\be
\label{ex}
H_\mathrm{sd}=-J \s_{i} \s_{\sigma\sigma'}\bb{S}_i\cdot \bb{\sigma}_{\sigma\sigma'}c^\dagger_{i\sigma}c\0_{i\sigma'},
\e
with the coupling strength $J$ taken here to be the same on $A$ and $B$ sublattices. The term (\ref{ex}) couples the tight-binding model for conduction electrons to a  classical Heisenberg model for localized angular momenta $\bb{S}_i$ with antiferromagnetic ground state. The localized momenta $\bb{S}_i$ are assumed to have large absolute value $|\bb{S}_i|=S\gg 1$. The characteristic $s$-$d$ exchange energy is given by $\Delta = J S$.  For conductivity and non-equilibrium spin density computations we assume the single-domain antiferromagnetic order, that is characterized by the unit N\'eel vector $\bb{\ell} = (\bb{S}^\textrm{A}-\bb{S}^\textrm{B})/2S$. 

Equal exchange couplings on both sublattices in Eq.~(\ref{ex}) and Rashba spin-orbit interaction of Eq.~(\ref{rashba}) ensure sublattice symmetry that greatly simplifies the results. In what follows we refer to the model as the symmetric model. At the end of the paper we also consider a more complex version of the model, where the sublattice symmetry is broken by a strong asymmetry of $s$-$d$ couplings on $A$ and $B$ sublattices.
 
\begin{figure}
\centering
\centerline{\includegraphics[width=\columnwidth]{{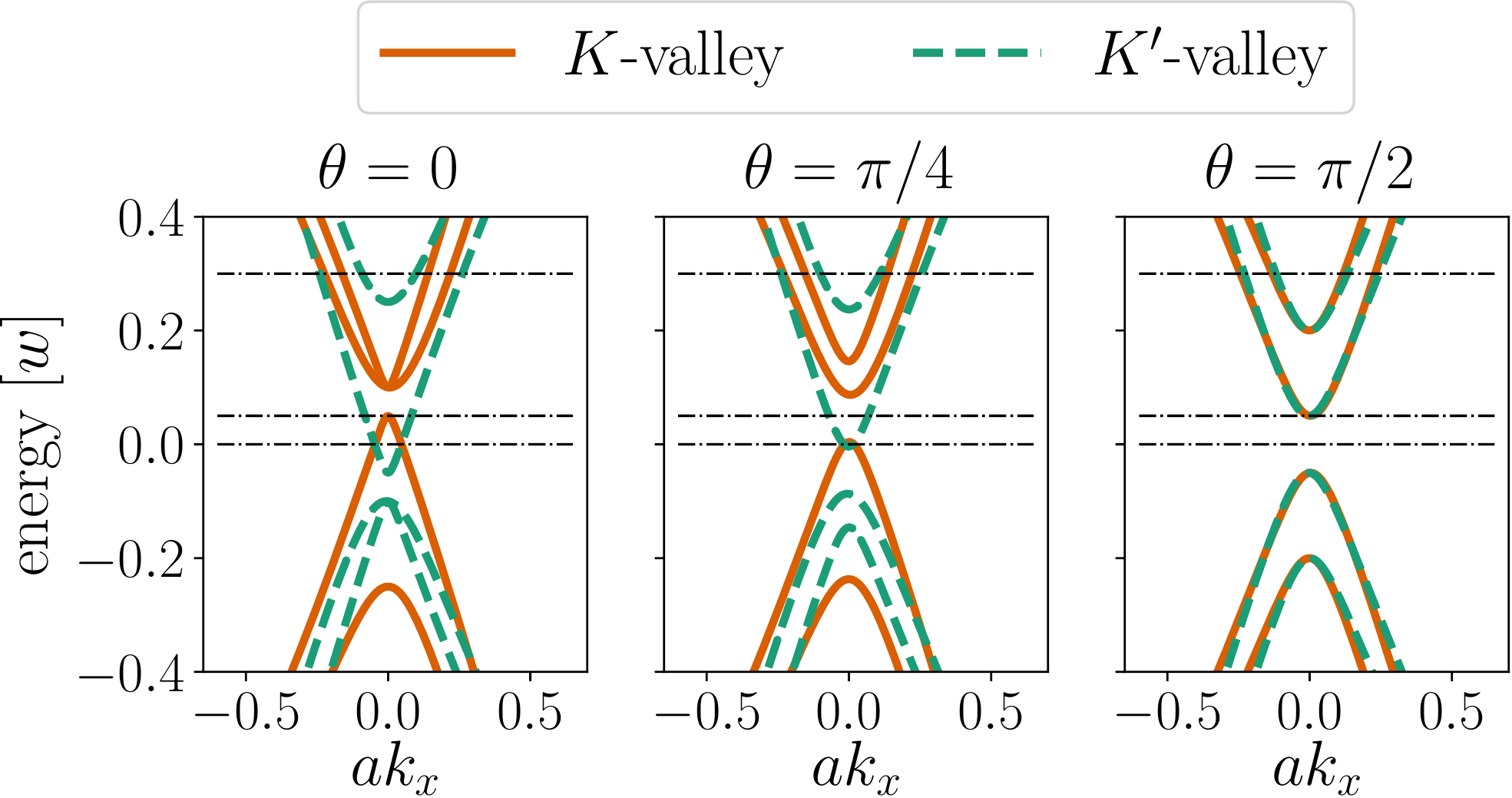}}}
\caption{
The band structure cross-section at $k_y=0$ for the model of Eq.~(\ref{ham}) versus $k_x$ counted with respect to the high-symmetry points $\bb{K}$ and $\bb{K}'$ for different orientations of the N\'eel vector, $\ell_z=\cos\theta$. Horizontal lines correspond to the Fermi energies $E = 0.0\,w$, $0.05\,w$, and $0.3\,w$ for which we perform the numerical analysis. For zero energy (band center), the system is insulating for $\theta>\pi/4$; for $\theta < \pi/4$ the Fermi energy crosses the conduction band in $K$ valley and the valence band in $K'$ valley. For $E=0.05\,w$ all extended states belong to the conduction band in $K$ valley that gives rise to 100\% valley-polarization. For $E=0.3\,w$ both valleys contain two Fermi surfaces. The plots correspond to the parameters $\lambda = 0.05\,w$ and $\Delta=0.1\,w$.
}
\label{fig:spectrum}
\end{figure}

The model of Eq.~(\ref{ham}) is motivated, in part, by the studies of CuMnAs \cite{SmejkalAFM_2017}. A similar model was also used to describe silicene where circularly polarized light induces a staggered $s$-$d$-interaction \cite{ezawa_photoinduced_2013}. The spin-orbit interaction of Eq.~(\ref{rashba}) breaks $\hat{\bb{z}}\to -\hat{\bb{z}}$ inversion symmetry. Similarly to CuMnAs we also assume that the Fermi energy of conduction electrons stay in a vicinity of the antiferromagnetic gap (or Dirac point). That means that the Fermi surfaces are located near two different pockets of the Brillouin zone -- the so-called $K$ and $K'$ valleys as shown in Fig.~\ref{fig:lattice}b (see also the caption in Fig.~\ref{fig:spectrum}). 

Due to the rather high symmetry of the model of Eqs.~(\ref{ham})-(\ref{ex}), the Fermi-surfaces remain entirely isotropic with respect to azimuthal (in-plane) direction of the N\'eel vector $\bb{\ell}$ as illustrated in Fig.~\ref{fig:lattice}b. Still, the band-structure of the symmetric model depends strongly on the polar angle $\theta$ of the N\'eel vector, $\ell_z= \cos\theta$, as shown in Fig.~\ref{fig:spectrum}.

\begin{figure}
\centering
\centerline{\includegraphics[width=\columnwidth]{{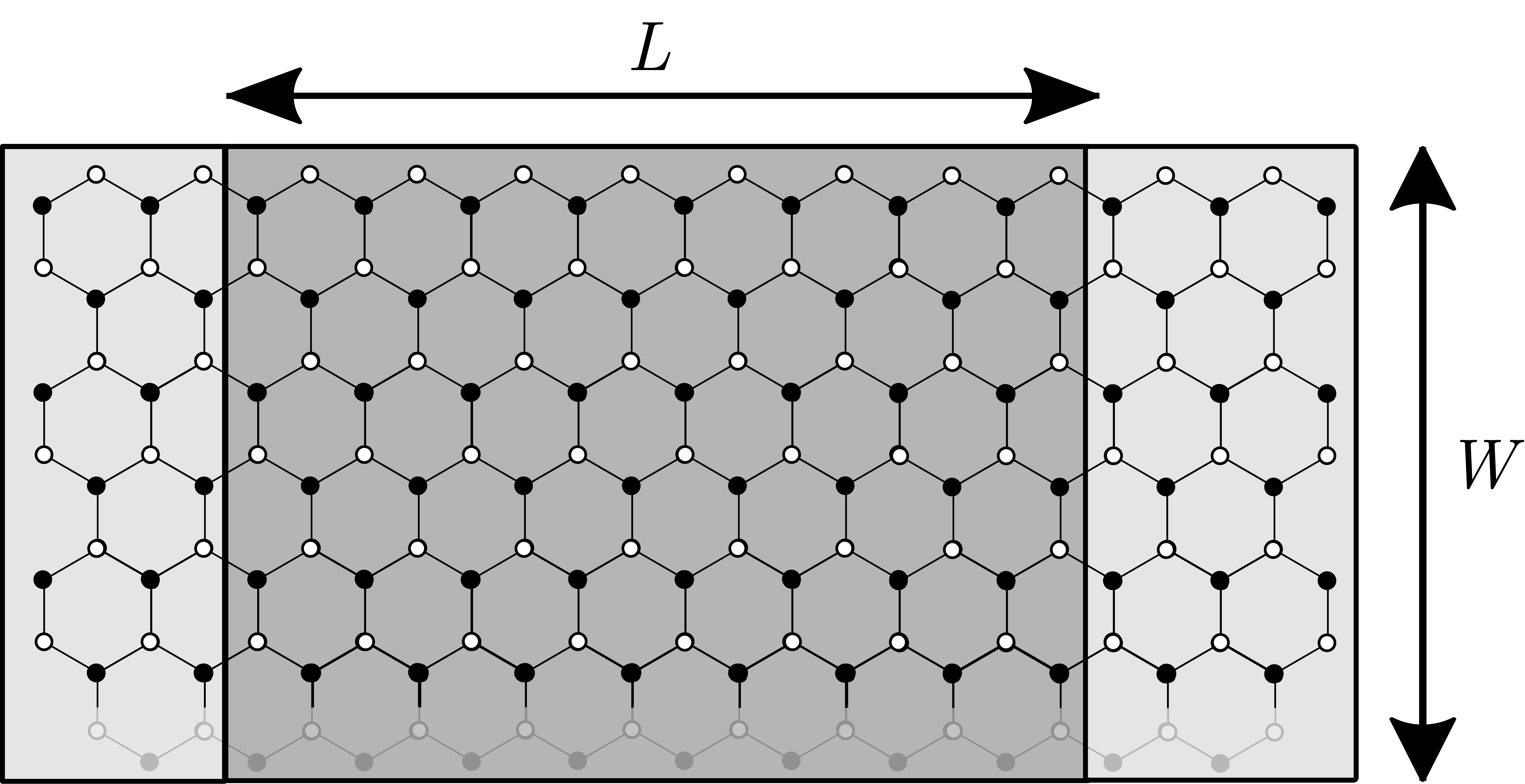}}}
\caption{Two-terminal geometry that is used for computation of spin-orbit torques. The largest sample corresponds to a wide zigzag nano-ribbon with $L \sim W \sim 750\,a$. We assume periodic boundary conditions in $y$ direction for both the sample and the leads. The sample is described by the model of Eq.~(\ref{MODEL}) with on-site disorder potential $V_i = \pm V_\textrm{d}$. Left and right leads ($x<0$ and $x>L$, respectively) are described by a clean Hamiltonian of Eq.~(\ref{ham}). The non-equilibrium spin density per current flux (averaged over the entire sample) is defined by the linear response formula of Eq.~(\ref{noneq}).}
\label{fig:sample}
\end{figure}

From the microscopic point of view all spin-orbit torques (as well as spin-transfer torques, spin-orbit induced Gilbert damping and effective renormalization of angular momenta $\bb{S}_i$) can be directly related to non-equilibrium contributions to the local spin density $\bb{s}_i$ that are proportional to electric current (or, in the case of Gilbert damping, to the time derivatives of the classical field $\bb{S}_i$) \cite{AdoSOT2017,AdoSTTGD2019}. The microscopic analysis of the torques is, therefore, similar to the microscopic analysis of the conductivity and must involve the mechanisms of momentum relaxation of conduction electrons that, in our system, is also directly related to the angular momentum relaxation of localized spins $\bb{S}_i$. Here we model such a momentum relaxation by adding non-magnetic on-site disorder potential to the model of Eq.~(\ref{ham}). Namely, we consider an ensemble of tight-binding models 
\be
\label{MODEL}
H=H_0+\s_{i}\s_{\sigma\sigma'}V_i\,c^\dagger_{i\sigma}c\0_{i\sigma'},
\e
where $V_i$ are random on-site potentials on randomly selected sites. For numerical simulations we take $V_i = \pm V_\textrm{d}$ (with a random sign) on random lattice sites in the scattering region (the sample) as illustrated in Fig.~\ref{fig:sample}. In various simulations we chose $V_\textrm{d} =0.2\,w$ and $V_\textrm{d} =0.5\,w$ and place impurities on 30\%, 40\% and 50\% of the lattice sites in the sample. Each point is obtained by averaging over 30 -- 80 disorder realizations. 

Even though the spin-orbit torques are completely defined by the tight-binding model of Eq.~(\ref{MODEL}), they are insufficient to describe magnetization dynamics of an antiferromagnet. The latter also depends on the type of the Heisenberg model used for $\bb{S}_i$ as well as on the Gilbert damping terms. In the presence of magnetic textures one should also take into account in-plane spin-transfer torques (which are defined by the response of spin density $\bb{s}_i$ to both electric current and spacial gradients of $\bb{S}_i$). Intimate relations between all these seemingly different (and often highly anisotropic) quantities have recently been established for a 2D Rashba ferromagnet in the metal regime \cite{AdoSTTGD2019}. 

\section{Scattering approach}

In this paper we present a numerical analysis of spin-orbit torques using the scattering framework. The framework appeals to the two terminal geometry depicted schematically in Fig.~\ref{fig:sample}. Left and right leads (reservoirs) are modeled by semi-infinite systems of the width $W$ described by ballistic tight-binding model of Eq.~(\ref{ham}) with periodic boundary conditions in $y$ direction. For both leads one constructs left and right-going scattering states, $\Psi^{L,\gtrless}_{\alpha,E}(\bb{r}_i)$ and $\Psi^{R,\gtrless}_{\alpha,E}(\bb{r}_i)$, that are the eigenstates of the model of Eq.~(\ref{ham}) that are normalized to the unit probability current flux through the lead cross-section. 

The scattering states are labeled by (i) the eigenenergy $E$; (ii) the lead index: $L$ for the left lead and $R$ for the right one; (iii) the flux direction: $>$ for the probability current in $x$ direction and $<$ for the probability current in the opposite ($-x$)  direction, and (iv) by a composite index $\alpha=(n,\sigma,\nu)$ that incorporates the channel index $n$ (numerating states with different projections of the wave vector on the transversal direction $y$), the spin projection $\sigma$ and the band index $\nu$ numerating Fermi surfaces. Note that the dimension of the scattering state wave-function is $1/\sqrt{Wv_\alpha}$, where $v_\alpha$ is the $x$-component of the velocity in the channel $\alpha$.  

With the help of the scattering states one can readily formulate a scattering problem at a given energy $E$ that is solved by the wave-function matching at the lead-sample interface for each disorder realization. For example, a scattering problem that corresponds to populating an incoming channel $\Psi^{L,>}_{\alpha,E}$ results in the eigenstate that has the following form in the leads 
\be
\label{scP1}
\Psi^L_{\alpha,E}(\bb{r})=\bc \Psi^{L,>}_{\alpha,E}(\bb{r})+\s_\beta r_{\beta\alpha} \Psi^{L,<}_{\beta,E}(\bb{r}),\quad &x<0,\\ \s_\beta t_{\beta\alpha} \Psi^{R,>}_{\beta,E}(\bb{r}),\qquad & x>L \ec, 
\e
where $t_{\beta\alpha}$ and $r_{\beta\alpha}$ denote the so-called transmission and reflection amplitudes, correspondingly. Similarly, the scattering problem that corresponds to populating a left-going state in the right lead, $\Psi^{R,<}_{\alpha,E}$, corresponds to the eigenstate
\be
\label{scP2}
\Psi^R_{\alpha,E}(\bb{r})=\bc \s_\beta t'_{\beta\alpha} \Psi^{L,<}_{\beta,E}(\bb{r}),\quad &x<0,\\  \Psi^{R,<}_{\alpha,E}(\bb{r})+\s_\beta r'_{\beta\alpha} \Psi^{R,>}_{\beta,E}(\bb{r}),\qquad & x>L \ec.
\e
The reflection and transmission amplitudes are organized into the scattering matrix 
\be
\label{scattering}
S=\bpm \hat{r} & \hat{t}' \\ \hat{t} & \hat{r}' \epm
\e
that yields the unitarity constraint $S\h S=1$. The constraint is ensured by the normalization of the scattering states $\Psi^{\gtrless}_{\alpha,E}(\bb{r})$ to the unit probability current flux that is conserved for each energy $E$ (in the absence of non-elastic processes).

The results of Eqs.~(\ref{scP1})-(\ref{scP2}) are routinely used to express, e.\,g. the time-averaged electric current flowing through the sample as
\be
\label{Icurrent}
I= e \int \frac{dE}{2\pi \hbar} \lt(f_R(E)-f_L(E)\rt) \s_{\alpha\beta} \lt|t_{\alpha\beta}(E)\rt|^2,
\e
where $f_L$ and $f_R$ stand for electron distribution functions in the left and right leads, respectively, and $e=-|e|$ is the electron charge. The expression for electric current leads to the celebrated \cite{datta_1995} Landauer-B\"uttiker formula for the conductance 
\be
\label{Landauer}
G=I/V_\textrm{bias} = \frac{e^2}{h} \s_{\alpha\beta} \lt|t_{\alpha\beta}(E)\rt|^2,
\e
where $V_\textrm{bias}$ is the voltage bias between the left and right leads (that sets a difference between the chemical potentials in the leads: $\delta\mu=\mu_L-\mu_R=e V_\textrm{bias}$) and $E$ is the average chemical potential in the sample (which we simply call the Fermi energy). 

The dependence of the average conductance on the sample length (for $L\simeq W$) is then fitted by the following formula
\be
\label{fit}
\la G \ra_\textrm{dis}= \frac{2e^2}{h}\, \frac{W}{L+\ell_0}\sigma,
\e 
where the angular brackets stand for averaging over disorder realizations, while the constant $\sigma$ is regarded as 2D dimensionless conductivity (which is independent of both $L$ and $W$). The length scale $\ell_0$ in Eq.~(\ref{fit}) is of the order of the transport mean free path in the system. 

\begin{figure}
\centering
\centerline{\includegraphics[width=\columnwidth]{{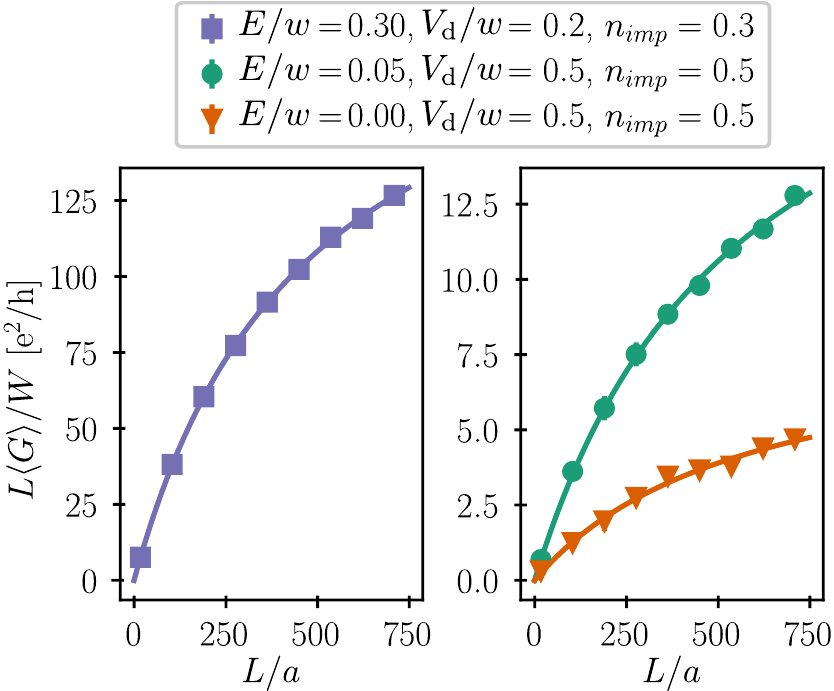}}}
\caption{Averaged two-terminal conductivity $L\la G\ra/W$ for the symmetric coupling model computed from Eq.~(\ref{Landauer}) as the function of the sample length $L$ for several values of the Fermi energy $E$, impurity strength $V_d$ and impurity concentration $n_\textrm{imp}$. Each point corresponds to averaging over 30 disorder realizations. The curves correspond to $\Delta=0.1\,w$, $\lambda=0.05\,w$ and $\theta=\pi/4$ and are fitted using Eq.~(\ref{fit}).  Similar behavior is observed for various angles $\theta$.}
\label{fig:conductance}
\end{figure}

\begin{figure}
\centering
\centerline{\includegraphics[width=\columnwidth]{{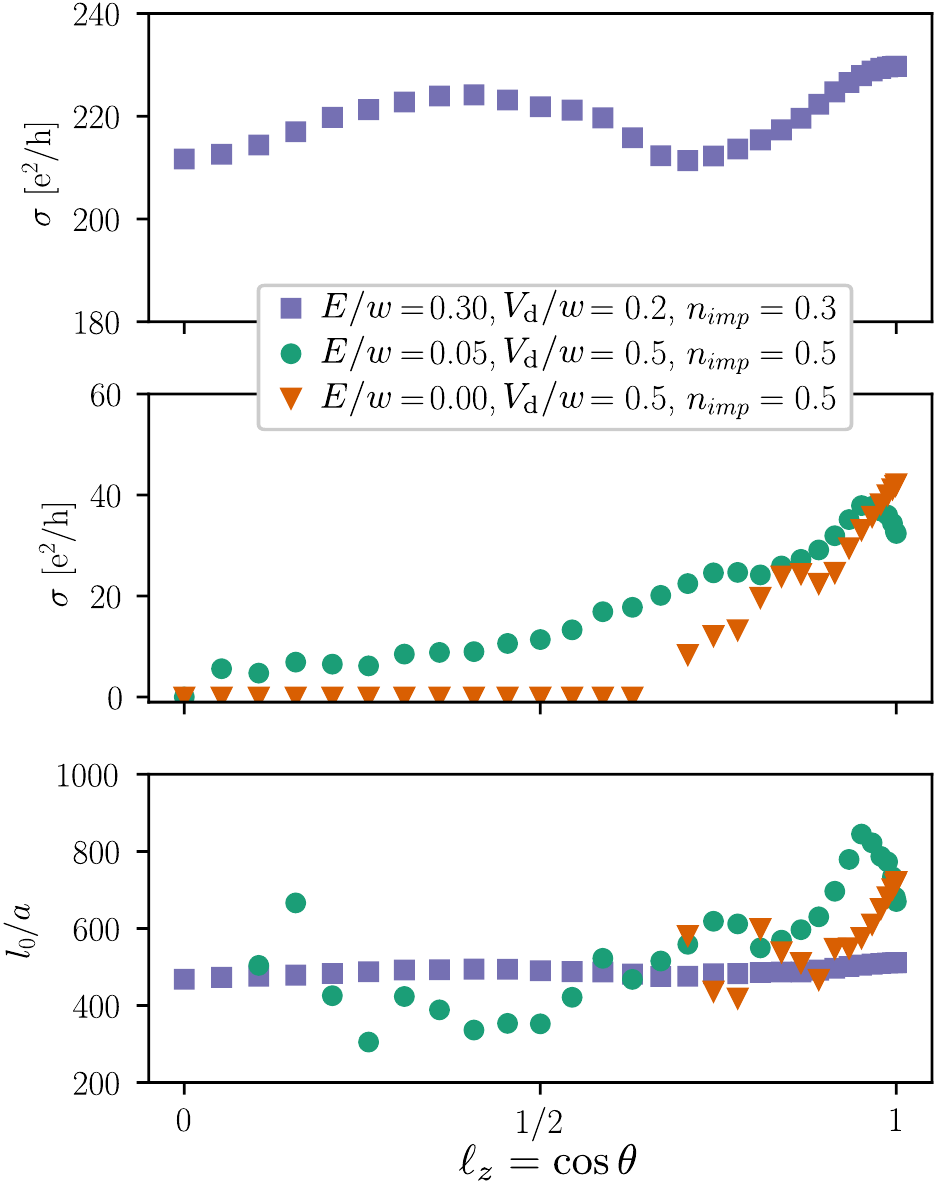}}}
\caption{The conductivity $\sigma$ and the mean free path $\ell_0$ in the symmetric coupling model of Eq.~(\ref{MODEL}) as the function of the N\'eel vector orientation $\ell_z=\cos\theta$. Both quantities are extracted from our numerical data with the help of  Eq.~(\ref{fit}). Large variations of the mean free path in the half-metal regime are associated with the Fermi energy touching the band edge at certain polar angles of the N\'eel vector.}
\label{fig:mfp}
\end{figure}

The straightforward numerical analysis using kwant package \cite{Groth2014} provides us with the two terminal conductivity, $L \la G\ra /W$, that is plotted in Fig.~\ref{fig:conductance} for the metal regime ($E=0.3\,w$, left panel) and for the half-metal regime ($E=0.05\,w$, right panel). The metal regime refers here to the Fermi energies that correspond to two Fermi surfaces per valley, while the half-metal regime corresponds to a single Fermi surface. Similar plots are generated for various polar angles $\theta$ of the N\'eel vector. Both the 2D conductivity $\sigma$ and the mean free path $\ell_0$ are, then, extracted from the fitting formula of Eq.~(\ref{fit}) and plotted in Fig.~\ref{fig:mfp} as the function of $\ell_z=\cos\theta$. The estimate of $\ell_0$ is necessary to ensure that our sample size is at least of the order of the mean free path to avoid non-universal finite-size effects in our data.  

One can see from Fig.~\ref{fig:mfp} that both the mean free path $\ell_0$ and the conductivity $\sigma$ are nearly constant in the metal regime. Their dependence on the N\'eel vector orientation is weak, which is consistent with the fact that such a dependence must disappear in the limit $E\gg \Delta$. In the half-metal regime, however, the conductivity is strongly dependent on both disorder concentration and the N\'eel vector angle $\theta$. We will see below, that despite rather strong angular dependence of the conductivity, one of the field-like spin-orbit torques remains $\theta$-independent in the half-metal regime. 

For $E=0$, the system exhibits metal-insulator transition as the function of the N\'eel vector orientation (see the second panel of Fig.~\ref{fig:conductance}). This choice of the Fermi energy leads, however, to vanishing spin-orbit torques as discussed below. 

\section{Spin-orbit torques}

The package kwant \cite{Groth2014} does not only provide the elements of the scattering matrix $S(E)$ in Eq.~(\ref{scattering}), but also gives access to the solutions $\Psi^{L,R}_{\alpha,E}(\bb{r})$ inside the sample. Such solutions can, therefore, be used to obtain a local expectation value of the electron spin (we refer here to the dimensionless spin defined by the operator $\bb{\sigma}/2$) as
\be
\label{sdensity}
\bb{s}(\bb{r})= \frac{1}{2}\int \frac{dE}{2\pi \hbar} \s_\textrm{A} f_A(E)  \sum_{\alpha}  \Psi^{A\;\dagger}_{\alpha,E} \bb{\sigma}_{\sigma\sigma'} \Psi^{A}_{\alpha,E}, 
 \e
where $\alpha=(n,\sigma,\nu)$ and the index $A=L,R$ numerates the leads. 

In order to extract spin-orbit torques we have to decompose the local electron spin to equilibrium and non-equilibrium contributions as
\be
\bb{s}(\bb{r})=\bb{s}_0(\bb{r})+\delta\bb{s}(\bb{r}).
\e
Equilibrium spin density, $\bb{s}_0$, corresponds to the limit of zero bias (or thermal bias) $f_L=f_R=f(E)$, where $f(E)$ is the Fermi-Dirac distribution function.  The quantity $\bb{s}_0$ describes equilibrium conduction electron contributions to the parameters of the Heisenberg model for localized momenta $\bb{S}_i$ that we do not discuss here. 

Non-equilibrium contribution, $\delta\bb{s}$, describes the effects induced e.\,g. by a bias voltage, $V_\textrm{bias}$, such as the spin-orbit torques on magnetization. In the linear response we obtain, in the spirit of Eq.~(\ref{Landauer}), the formula
\begin{align}
\frac{\delta\bb{s}(\bb{r})}{\delta\mu} =  \frac{1}{4h} \sum_{\alpha}  &\lt[\Psi^{R\;\dagger}_{\alpha,E}(\bb{r}) \bb{\sigma}_{\sigma\sigma'} \Psi^{R}_{\alpha,E}(\bb{r}) \rt. \n\\
&\; \lt. - \Psi^{L\;\dagger}_{\alpha,E}(\bb{r}) \bb{\sigma}_{\sigma\sigma'} \Psi^{L}_{\alpha,E}(\bb{r}) \rt],
\label{noneq}
\end{align}
where $\delta\mu=eV_\textrm{bias}$ and the scattering states are taken at the Fermi energy.  Similarly to Eq.~(\ref{Landauer}), we assume here that the quantities $\Psi^{A\;\dagger}_{\alpha,E} \,\bb{\sigma}\, \Psi^{A}_{\alpha,E}$  have negligible energy dependence within the interval $\delta\mu$ around the Fermi energy. 

In numerical simulations we systematically employ Eq.~(\ref{noneq}) to extract the spin density on $A$ and $B$ sublattices by varying the direction of the N\'eel vector. The results are fitted by the expansion of $\delta\bb{s}$ in angular harmonics that are compatible with the symmetry of the model. Before, we proceed with the discussion of our results let us pause to clarify the relation between the non-equilibrium spin-density and microscopic spin-orbit torques on magnetization that enter effective Landau-Lifshits-Gilbert (LLG) equation on magnetization dynamics.

\section{Microscopic LLG equation} 

The problem of magnetization dynamics can be formulated with the help of a Heisenberg model for the localized momenta $\bb{S}_i$ that is coupled to a tight-binding model for conduction electrons by means of the local exchange interaction of Eq.~(\ref{ex}). Depending on the type of the Heisenberg model we may obtain different equations of motion but the torques on magnetization (originating from electric current or field) and also Gilbert damping terms can be understood from the tight-binding Hamiltonian of Eq.~(\ref{MODEL}) alone \cite{AdoSTTGD2019}. 

For the sake of illustration let us outline the derivation of magnetization dynamics for a Heisenberg model of an ``isotropic'' antiferromagnet. The local interaction of Eq.~(\ref{ex}) dictates the equation of motion of localized momenta of the form
\be
\label{EOM}
\dot{\bb{S}}_i = \bb{H}_i\times \bb{S}_i+ \frac{J}{\hbar}\, \bb{S}_i\times \s_{\sigma\sigma'} \lt\la c\h_{i\sigma}\bb{\sigma}_{\sigma\sigma'} c\0_{i\sigma'} \rt\ra,
\e
where dot stands for the time derivative of $\bb{S}_i$ and $\bb{H}_i=-\delta F/ \delta\, \hbar\bb{S}_i$ is an effective field (in frequency units) on a site $i$  that is defined by the free energy $F$ of the classical Heisenberg model for localized spins. The angular brackets denote the thermodynamic expectation value of conduction electron spin operator on the same site $i$. 

For a collinear antiferromagnet we shall distinguish two sublattices $A$ and $B$ that are characterized by opposite directions of the localized momenta $\bb{S}_i^\textrm{A}=-\bb{S}_i^\textrm{B}$ in the antiferromagnetic ground state. It is assumed, that these directions remain to be almost opposite even for out-of-equilibrium conditions. It is also natural to assume that, as far as the magnet is far from a phase transition, the classical fields $\bb{S}_i^\textrm{A}$ and $\bb{S}_i^\textrm{B}$ remain smooth on atomic scales. Below we consider a single domain antiferromagnet, which is characterized by two time-dependent unit vectors $\bb{n}^\textrm{A,B}=\bb{S}^\textrm{A,B}/S$. 

It is also convenient to define smooth conduction electron spin densities on each of the two sublattices
\be
\bb{s}^\textrm{A,B}(\bb{r})= \frac{1}{2} \s_{i\sigma\sigma'} \lt\la c\h_{i\sigma}\bb{\sigma}_{\sigma\sigma'} c\0_{i\sigma'} \rt\ra\;\frac{2}{\mathcal{A}},
\e
where $\mathcal{A}$ is the area of the unit cell (which naturally includes one A and one B site). Thus, the equation of motion (\ref{EOM}) can be written in  continuous approximation as
\beml
\label{EOMAFM}
\begin{align}
\dot{\bb{n}}^\textrm{A} &= \bb{H}^\textrm{A}\times \bb{n}^\textrm{A}+ (J\mathcal{A}/\hbar)\, \bb{n}^\textrm{A}\times \bb{s}^\textrm{A},\\
\dot{\bb{n}}^\textrm{B}&= \bb{H}^\textrm{B}\times \bb{n}^\textrm{B}+ (J\mathcal{A}/\hbar)\, \bb{n}^\textrm{B}\times \bb{s}^\textrm{B},
\end{align}
\eml
where $|\bb{n}^\textrm{A,B}|=1$, and the notations $\bb{H}^\textrm{A,B}$ refer to the effective fields on the sublattices $A$ and $B$. 

Antiferromagnet dynamics is usually formulated as the coupled dynamics of the N\'eel and magnetization vectors,
\be
\bb{\ell}=\lt(\bb{n}^\textrm{A}-\bb{n}^\textrm{B}\rt)/2,\qquad \bb{m}= \lt(\bb{n}^\textrm{A}+\bb{n}^\textrm{B}\rt)/2,
\e
that remain mutually perpendicular $\bb{\ell}\cdot \bb{m}=0$ and yield the constraint $\ell^2+m^2=1$. Naturally, the amplitude of the magnetization vector remain to be small, $m\ll \ell \approx 1$.

For an ``isotropic'' antiferromagnet, one finds the effective field \cite{Gomonay2014} $\bb{H}^\textrm{A}+\bb{H}^\textrm{B}= J_\textrm{ex}\bb{m}/\hbar+2\bb{H}$, where $\bb{H}$ is an external magnetic field in frequency units and $J_\textrm{ex}$ is a direct antiferromagnetic exchange energy that is one of the largest energies in the problem. In turn, the combination $\bb{H}^\textrm{A}-\bb{H}^\textrm{B}$ is proportional to magnetic anisotropy that we choose not to take into account as mentioned above.  

Equations (\ref{EOMAFM}) can, therefore, be rewritten as
\begin{align}
&\dot{\bb{\ell}} = -\frac{J_\textrm{ex}}{2\hbar} \bb{\ell}\times\bb{m} +\bb{H}\times\bb{\ell}+\frac{J\mathcal{A}}{\hbar}\lt(\bb{\ell}\times\bb{s}_++\bb{m}\times\bb{s}_-\rt),\n\\
&\dot{\bb{m}} = \bb{H}\times\bb{m}+\frac{J\mathcal{A}}{\hbar}\lt(\bb{m}\times\bb{s}_++\bb{\ell}\times\bb{s}_-\rt),
\label{AFMEOM}
\end{align}
where $\bb{s}_\pm=(\bb{s}_A\pm\bb{s}_B)/2$. The right-hand sides of Eqs.~(\ref{AFMEOM}) contain the quantities that can be called generalized torques. Here we are only interested in specific contributions to generalized torques that are induced by the chemical potential difference $\delta\mu=eV_\textrm{bias}$ between left and right leads. Such contributions define four spin-orbit torques  
\be
\label{SOT_def}
\bb{T}^\ell_\pm=(J\mathcal{A}/\hbar)\bb{\ell}\times\delta\bb{s}_\pm, \quad 
\bb{T}^m_\pm=(J\mathcal{A}/\hbar)\bb{m}\times\delta\bb{s}_\pm,
\e
where $\delta \bb{s}_\pm$ refers to the non-equilibrium spin density contribution that is proportional to $\delta\mu$. 

Note that, generally, the average spin $\bb{s}_i$ has a non-local functional dependence on the time-dependent classical field $\bb{S}_j$ at preceding moments of time  and at different lattice sites $j\neq i$. The degree of non-locality is defined by relaxation processes. The quicker the relaxation the more local the functional dependence. In particular, non-dissipative contributions to spin-orbit torques (the so-called field-like contributions) are local in time on the time scales of the order of $s$-$d$ exchange  $\tau_\textrm{sd}=\hbar/\Delta$, where $\Delta=J S$. In contrast, the dissipative contributions (such as anti-damping torques) are defined by transport scattering time $\tau_\textrm{tr}$. The latter time scale may be both larger and smaller than $\tau_\textrm{sd}$ giving rise to different regimes of magnetization dynamics. For the sake of numerical analysis we consider, however, a situation when these two time scales are of the same order.

\section{Symmetry consideration} 

Symmetry properties of spin-orbit torques and conductivity tensor can be understood from a low-energy approximation by projecting the model (\ref{ham}) on the states in a vicinity of $K$ points, 
\be
\bb{K}= \frac{4\pi}{3\sqrt{3} a}\bpm 1\\ 0\epm,\quad\mbox{and}\quad \bb{K}'= -\bb{K},
\e
as it is usually done for graphene. In the valley symmetric approximation we obtain the effective model
\be
\label{low}
H^\textrm{eff}_0= v\, \bb{p}\cdot\bb{\Sigma}+\alpha_\textrm{R}\lt[\bb{\sigma}\times\bb{\Sigma}\rt]_{\hat{z}} - \Delta\,\bb{\ell}\cdot\bb{\sigma}\,\Sigma_z\Lambda_z,
\e
where $\bb{\Sigma}$, $\bb{\Lambda}$, and $\bb{\sigma}$ are the vectors of Pauli matrices in sublattice, valley and spin space, respectively, and 
\be
v= \frac{3}{2} wa/\hbar, \qquad \alpha_\textrm{R}= \frac{3}{2} \lambda , \qquad \Delta=JS.
\e
Even though the model of Eq.~(\ref{ham}) is characterized by the point group $C_\textrm{3v}$, the effective low-energy model of Eq.~(\ref{low}) has a higher symmetry 
of the group $C_{\infty\textrm{v}}$. This is reflected by the fact that the Fermi-surfaces (for energies we are interested in) remain entirely isotropic irrespective of the N\'eel vector orientation as shown in Fig.~\ref{fig:lattice}. Moreover, the exact sublattice symmetry of the model, $\Lambda_x H^\textrm{eff}_0[-\bb{\ell}] \Lambda_x =H^\textrm{eff}_0[\bb{\ell}]$, ensures vanishing non-equilibrium staggered spin polarization $\delta \bb{s}_-=0$ irrespective of both the voltage bias and the N\'eel vector orientation. This property is indeed confirmed by the numerical analysis. 

The spin-orbit coupling term in Eq.~(\ref{low}) is proportional to the scalar product $\bb{\sigma}\cdot \lt[\hat{\bb{z}}\times \bb{v}\rt]$, where $\bb{v}=v\bb{\Sigma}$ is the velocity operator and $\hat{\bb{z}}$ is the unit vector in $z$ direction. In the presence of electric current $I$, which, in our model, is always directed along $x$ axis, the velocity operator averages to a non-zero value that is proportional to the current. One can see, therefore, that the second term in Eq.~(\ref{low}) can be interpreted as Zeeman coupling to an effective Rashba field $\hat{\bb{z}}\times \bb{I}$. Such a field induces non-equilibrium polarization $\delta\bb{s}_+$ in the same way as in ferromagnets. 

The symmetry of the point group $C_{\infty\textrm{v}}$ (see also the symmetry analysis in Refs.~\onlinecite{Duine2012, Garello2013}) suggests that the tensor relation between non-equilibrium polarization $\delta\bb{s}_+$ and the Rashba field $\hat{\bb{z}}\times \bb{I}$ can be presented in a vector form as 
\begin{align}
\delta\bb{s}_+&\,= A_I(\ell_z^2)\,\hat{\bb{z}}\times \bb{I} + A'_I(\ell_z^2)\,\bb{\ell}_\parallel\times \lt[\bb{\ell}_\parallel\times \lt[\hat{\bb{z}}\times \bb{I}\rt]\rt]\n\\
&+ B_\perp(\ell_z^2)\,\bb{\ell}_\perp\times \lt[\hat{\bb{z}}\times \bb{I}\rt]+B_\parallel(\ell_z^2)\,\bb{\ell}_\parallel\times \lt[\hat{\bb{z}}\times \bb{I}\rt]\n\\
&+ C(\ell_z^2)\,\bb{\ell}_\parallel\times \lt[\bb{\ell}_\perp\times \lt[\hat{\bb{z}}\times \bb{I}\rt]\rt],
\label{symmetry}
\end{align}
where we decompose the N\'eel vector $\bb{\ell}=\bb{\ell}_\parallel+\bb{\ell}_\perp$ into in-plane $\bb{\ell}_\parallel$ and perpendicular-to-the-plane $\bb{\ell}_\perp$ components. We parameterize $\bb{\ell}=(\cos\phi\sin\theta,\sin\phi\sin\theta,\cos\theta)^\top$, where $\theta$ is a polar angle and $\phi$ is an azimuthal angle. Consequently, we have $\bb{\ell}_\parallel=(\cos\phi\sin\theta,\sin\phi\sin\theta,0)^\top$ and $\bb{\ell}_\perp=(0,0,\cos\theta)^\top$.

The decomposition of Eq.~(\ref{symmetry}) is dictated by the symmetry analysis with respect to two transformations: the N\'eel vector inversion, $\bb{\ell}\to -\bb{\ell}$, and the N\'eel vector reflection, $\bb{\ell}_\perp\to -\bb{\ell}_\perp$, with respect to the electron 2D plane. 

Since the model (at the Fermi energies we choose) is isotropic with respect to in-plane rotations, these two symmetries are the only ones to characterize different contributions to non-equilibrium spin density. 

The first two terms in Eq.~(\ref{symmetry}) are even with respect to both the N\'eel vector inversion and the N\'eel vector reflection. The last term in Eq.~(\ref{symmetry}) is even with respect to the N\'eel vector inversion but odd with respect to the N\'eel vector reflection. These terms correspond to torques of the field-like symmetry, i.\,e. to the torques that are invariant under the time reversal operation. 

The second two terms in Eq.~(\ref{symmetry}), which are proportional to the coefficients $B_{\parallel,\perp}$, are odd with respect to the N\'eel vector reflection. The first one of them is odd while the second one is even with respect to the N\'eel vector reflection. These terms represent anti-damping spin-orbit torques that change sign under time reversal.

The coefficients in front of different vector forms in Eq.~(\ref{symmetry}) must remain invariant with respect to both symmetry transformations, but they may still vary as the function of the symmetry invariant $\ell_z^2$. The coefficients may also depend on $\ell^2$ but such dependence is absent within linear response approximation. Indeed, one shall take $\ell^2=1$ ($m=0$) in all coefficients in front of electric current since electric current is regarded as the only possible source of a deviation from the exact antiferromagnetic order. 

The exact sublattice symmetry of the symmetric coupling model of Eqs.~(\ref{ham})-(\ref{MODEL}) does not only ensure the vanishing staggered polarization $\delta\bb{s}_-=0$ but also leads to identically vanishing anti-damping torques $B_\parallel=B_\perp=0$ in all possible regimes. This is indeed confirmed by our numerical analysis for the symmetric coupling model as we describe below. 

\begin{figure}
\centering
\centerline{\includegraphics[width=\columnwidth]{{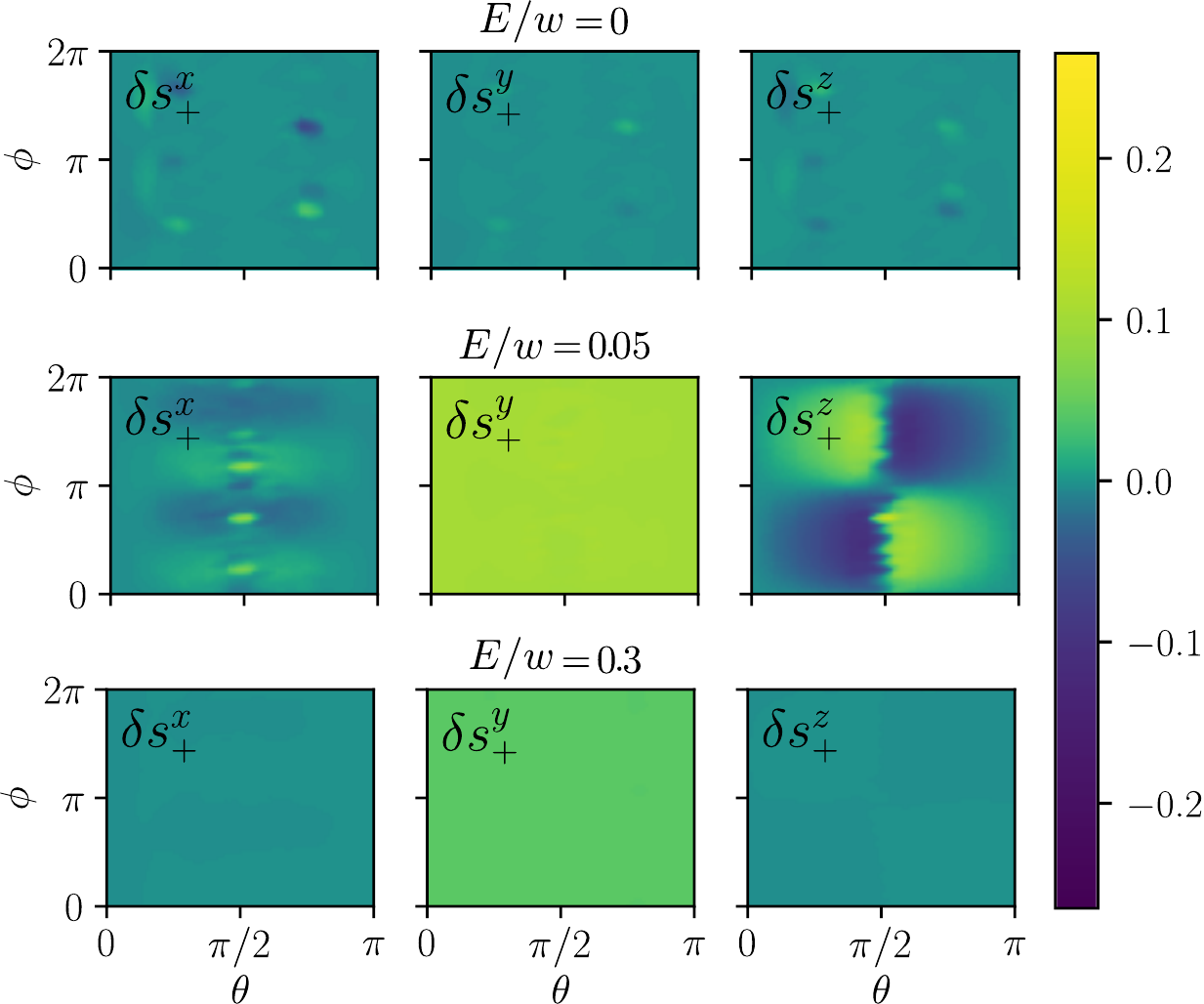}}}
\caption{Non-equilibrium spin density $\delta\bb{s}_+$ for the symmetric coupling model of Eqs.~(\ref{ham})-(\ref{MODEL}) with $V_\textrm{d}=0.5\,w$ as the function of the N\'eel-vector orientation for different values of the Fermi energy. Each data point foe each direction of the N\'eel vector corresponds to averaging over 30 impurity configurations. Spin densities are normalized to the scale $\Delta\lambda j/evw^2$, where $j=I/W$ is the charge current density through the sample. Fluctuations at $\theta=\pi/4$ and $\theta=3\pi/4$ in the top panel and at $\theta=\pi/2$ in the middle panel are due to diverging mean free path for Fermi energy touching the bottom of the conduction band. The staggered spin density $\delta\bb{s}_-$ fluctuates around zero (not-shown). For $E/t=0$, the spin-density $\delta\bb{s}_+$ also fluctuates around zero as shown in the upper panel.}
\label{fig:density}
\end{figure}

\section{Results for symmetric model} 
\subsection{General}

With the help of the kwant package \cite{Groth2014} we compute from Eq.~(\ref{noneq}) the mean spin density response $\delta\bb{s}/\delta\mu$ on both sublattices for various directions of the N\'eel vector $\bb{\ell}$ and three different Fermi energies. These results, which are presented in Fig.~\ref{fig:density}, have complex dependence on the orientation of the N\'eel vector. The dependence is, however, much simplified for non-equilibrium spin density response divided by the sample conductance. In other words, we observe that the non-equilibrium spin density has a much simpler form when expressed via the charge current density $\bb{j}=\bb{I}/W$ rather than via the bias voltage.

In all regimes considered we find $\delta\bb{s}_-=0$ and confirm the decomposition of Eq.~(\ref{symmetry}) with $B_\parallel=B_\perp=0$. Moreover, we also find that the coefficient $A_I(\ell_z^2)=A_I$ is constant, i.\,e. independent of the angle $\theta$ in all regimes we consider. In addition, we find that the non-equilibrium spin density vanishes identically, $\delta\bb{s}_+=\delta\bb{s}_-=0$ for $E=0$. This point corresponds to the exact electron-hole symmetry of the model, i.\,e. for $E=0$, the number of quasiparticles in the conduction and valence bands are exactly the same. 

Despite these general findings we still observe that the results for non-equilibrium spin density are qualitatively different in the metal and half-metal regimes. 

\begin{figure}
\centering
\centerline{\includegraphics[width=\columnwidth]{{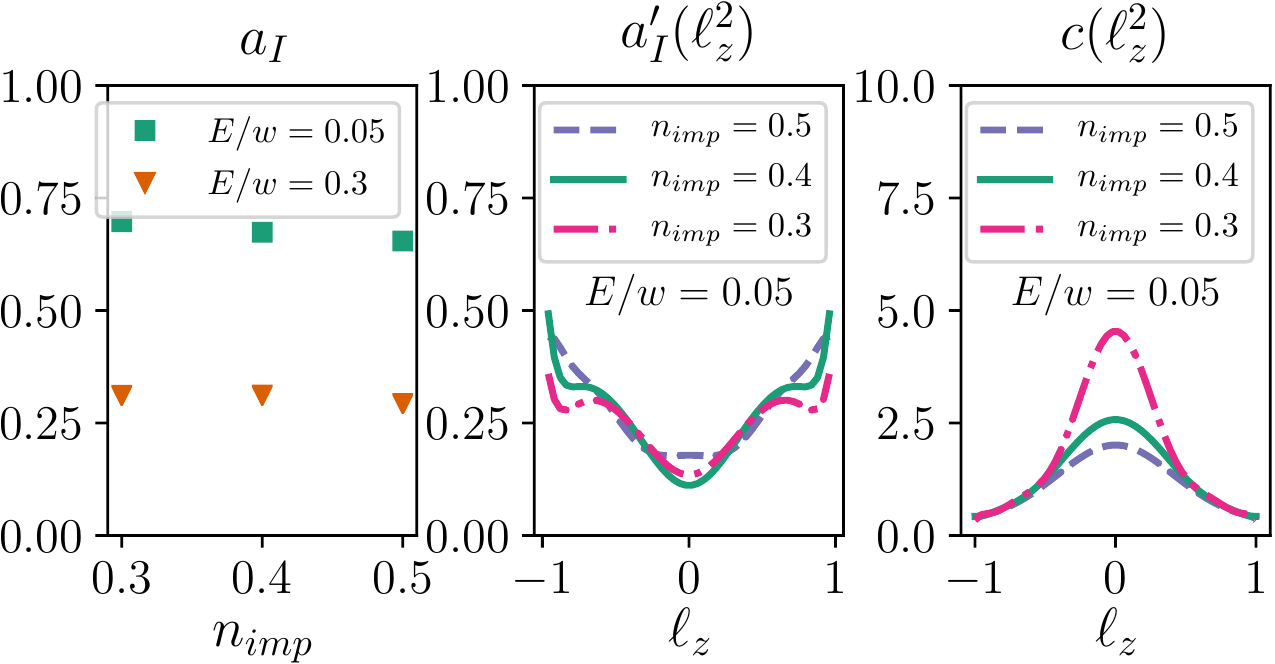}}}
\caption{The results of fitting simulation data for the symmetric coupling model with $L=2W=750\,a$ (partly shown in Fig.~\ref{fig:density}) with the ansatz of Eq.~(\ref{result1}). The data is obtained for $V_\textrm{d}=0.5\,w$. The data have been collected for 200 different orientations of the N\'eel vector for each impurity concentration.}
\label{fig:decomposition}
\end{figure}

\subsection{Metal regime}

The metal regime is characterized by two Fermi surfaces for each of the two valleys. This regime is represented in our simulations by the choice $E=0.3\,w$. In this case we also find $A'_I=C=0$ within our numerical accuracy. Thus, the metal regime in the symmetric model is represented by the only contribution to the non-equilibrium spin-density
\be
\label{s_SOT}
\delta\bb{s}_+= \frac{\Delta\lambda}{2\pi evw^2} a_I\,\hat{\bb{z}}\times \bb{j}, 
\e
with a constant dimensionless coefficient $a_I$. This coefficient retains almost the same value for three different impurity concentrations as 
shown in the left panel of Fig.~\ref{fig:decomposition}. Thus, the non-equilibrium spin density in the metal regime is simply proportional to the Rashba field. The non-equilibrium spin density of Eq.~(\ref{s_SOT}) is readily recognized as the inverse spin-galvanic effect of Edelstein \cite{Edelstein1990} that is widely known in ferromagnet materials with Rashba coupling \cite{ManchonPRB2008, GarateSOT2009}. 

Consequently, the spin orbit torques in the metal regime of the symmetric model are given solely by the isotropic Edelstein effect as
\be
\label{metal_SOT}
\bb{T}^\ell_+=a_I \eta\;\bb{\ell} \times\lt[\hat{\bb{z}}\times\bb{j}\rt],\quad
\bb{T}^m_+=a_I \eta\;\bb{m} \!\times\!\lt[\hat{\bb{z}}\times\bb{j}\rt], 
\e
where $\eta=J\mathcal{A}\Delta\lambda /evh w^2$. 

This results of Eqs.~(\ref{s_SOT}), (\ref{metal_SOT}) are remarkably similar to those found analytically in the metal regime of a Rashba ferromagnet model with white-noise disorder \cite{AdoSOT2017,AdoSTTGD2019}. The latter is also characterized by identically vanishing anti-damping spin-orbit torques and by a completely isotropic field-like torque of the same symmetry. Even though, analytical results have been obtained for a ferromagnet with a weak white-noise disorder, the same drastic simplification of the spin-orbit torque takes place for a symmetric honeycomb antiferromagnet with rather strong point-like disorder as demonstrated above. Such a simplification is, however, limited to the metal regime. 
 
\subsection{Half-metal regime} 
 
The half-metal regime is represented by the Fermi energy $E=0.05\,w$ that corresponds to the presence of a single Fermi surface in one of the valleys. The conductivity in this regime acquires strong dependence on $\theta$. In particular, for $\ell_z^2<1/2$, the system is poorly conducting as one can see already in the second panel of Fig.~\ref{fig:mfp}. Remarkably, we still find that the coefficient $A_I$ in Eq.~(\ref{symmetry}) is entirely independent of $\theta$ and anti-damping like torques are vanishing identically, $B_{\parallel,\perp}=0$. The coefficients $A'_I$ and $C$ are, however, finite and acquire some dependence on $\ell_z^2$.   

Thus, our numerical data in half-metal regime corresponds to the spin density
\begin{align}
\label{result1}
\delta\bb{s}_+= \frac{\Delta\lambda}{2\pi evw^2}
\Big(\,&a_I\,\hat{\bb{z}}\times \bb{j} + a'_I(\ell_z^2)\,\bb{\ell}_\parallel\times \lt[\bb{\ell}_\parallel\times \lt[\hat{\bb{z}}\times\bb{j}\rt]\rt]\n\\
&+c(\ell_z^2)\,\bb{\ell}_\parallel\times \lt[\bb{\ell}_\perp\times \lt[\hat{\bb{z}}\times\bb{j}\rt]\rt]\Big),
\end{align}
where we also introduced the dimensionless functions $a'_I(\ell_z^2)$ and $c(\ell_z^2)$ that are shown in the middle and in the right panel of Fig.~\ref{fig:decomposition} for different impurity concentrations. 

The last two terms in Eq.~(\ref{result1}) represent high-harmonics field-like torques that are finite only in the half-metal regime. On symmetry grounds one should generally expect the field-like torques to be largely insensitive to disorder. Even though, the disorder dependence of $a_I$ and $a'_I$ coefficients is indeed negligible, the one of the function $c(\ell_z^2)$ is still rather strong for $\ell_z^2<1/2$. One should, however, remember that the coefficient $c$ is standing in front of the vector form that is vanishing for both $\ell_z=0$ and $\ell_z=\pm 1$, so the fit accuracy near these points is poor. Moreover, the case of almost in-plane N\'eel vector, $\ell_z^2\ll 1/2$, corresponds to the poorly conducting sample whose conductance is dominated by the variable range hopping processes due to strong disorder. Thus, we attribute this seemingly strong dependence of $c$ on disorder concentration to the mechanism of conduction. Still, the value of $c$ is found to be about 5 times larger than that of $a_I$ and about 10 times larger than that of $a'_I$, which makes the high-harmonic $c$-torque relevant in the half-metal regime. We note that the non-equilibrium spin density, which is proportional to $c(\ell_z^2)$, is directed perpendicular to the electron plane.  

Overall, the half-metal regime is characterized by the torques
\beml
\label{torque_T}
\begin{align}
\bb{T}^\ell_+=\eta \Big(\,&a_I\bb{\ell} \!\times\!\lt[\hat{\bb{z}}\!\times\!\bb{j}\rt] + 
a'_I(\ell_z^2)\, \bb{\ell}\!\times\!\lt[\bb{\ell}_\parallel\!\times\! \lt[\bb{\ell}_\parallel\!\times\! \lt[\hat{\bb{z}}\!\times\! \bb{j}\rt]\rt]\rt]\n\\
&+c(\ell_z^2)\, \bb{\ell}\!\times\!\lt[\bb{\ell}_\parallel\!\times\! \lt[\bb{\ell}_\perp\!\times\! \lt[\hat{\bb{z}}\!\times\! \bb{j}\rt]\rt]\rt]\Big),
\label{torque_Tplus}\\
\bb{T}^m_+=\eta \Big(\,&a_I\bb{m} \!\times\!\lt[\hat{\bb{z}}\!\times\!\bb{j}\rt] + 
a'_I(\ell_z^2)\, \bb{m}\!\times\!\lt[\bb{\ell}_\parallel\!\times\! \lt[\bb{\ell}_\parallel\!\times\! \lt[\hat{\bb{z}}\!\times\! \bb{j}\rt]\rt]\rt]\n\\
&+c(\ell_z^2)\, \bb{m}\!\times\!\lt[\bb{\ell}_\parallel\!\times\! \lt[\bb{\ell}_\perp\!\times\! \lt[\hat{\bb{z}}\!\times\! \bb{j}\rt]\rt]\rt]\Big),
\label{torque_Tminus}
\end{align}
\eml
where $\eta=J\mathcal{A}\Delta\lambda /evh w^2$ is the same as in Eq.~(\ref{metal_SOT}). 

Note, that the function $a'_I(\ell_z^2)$ in Eq.~(\ref{torque_Tplus}) for $\bb{T}^\ell_+$ can be absorbed into the redefinition of the coefficients $a_I$ and $c$.  Indeed, with the help of a straightforward vector algebra one can re-write Eq.~(\ref{torque_Tplus}) as
\be
\label{torque_Tplus_MOD}
\bb{T}^\ell_+=\eta \Big(\tilde{a}_I(\ell_z^2)\bb{\ell} \!\times\!\lt[\hat{\bb{z}}\!\times\!\bb{j}\rt] 
+\tilde{c}(\ell_z^2)\, \bb{\ell}\!\times\!\lt[\bb{\ell}_\parallel\!\times\! \lt[\bb{\ell}_\perp\!\times\! \lt[\hat{\bb{z}}\!\times\! \bb{j}\rt]\rt]\rt]\Big),
\e
where we introduced
\be
\label{renorm}
\tilde{a}_I=a_I-(1-\ell_z^2)a'_I, \qquad \tilde{c}=c-a'_I.
\e
Thus, the modified coefficient $\tilde{a}_I$ acquires a weak dependence on $\ell_z^2$ due to a finite, though small, $a'_I$ contribution in the half-metal regime. 

To conclude this section it is worth mentioning that half-metal antiferromagnets are not entirely hypothetical. There have been several proposals in the past (see Refs.~\onlinecite{deGroot1995, Gong2018} to name a few) and there has been also a recent material Mn$_2$Ru$_\textrm{x}$Ga that is recognized as a half-metal antiferromagnet \cite{Kurt2014, Coey2016}.

\section{Asymmetric coupling model} 

Relative simplicity of the results of Eqs.~(\ref{metal_SOT},\ref{torque_T}) for the symmetric coupling model of Eqs.~(\ref{ham})-(\ref{MODEL}) prompted us to look at a model with ultimately asymmetric $s$-$d$ exchange,
\be
H_\mathrm{sd}=\,-J \s_{i\in A }  \s_{\sigma\sigma'} \bb{\sigma}_{\sigma\sigma'}\bb{S}^\textrm{A}_i\cdot c^\dagger_{i\sigma}c\0_{i\sigma'},
\label{ex2}
\e
that is represented by $s$-$d$ coupling on $A$ sublattice only. Such a coupling obviously violates the sublattice symmetry and may, in principle, lead to the appearance of non-equilibrium staggered polarization and to anti-damping like spin-orbit torques that are absent in the symmetric coupling model. 

It is worth mentioning that the asymmetric model is unlikely to represent an antiferromagnet. Since conduction electron spins contribute to polarization only on a single sublattice, the model is much more likely to correspond to a ferrimagnet or a ferromagnet.  Rare-earth/transition metal ferrimagnets (such as FeCoGd) at a compensation point seem to be the most obvious examples \cite{Hoffman2018}. The conduction electrons in these compounds couple mostly with localized d-orbitals rather than with f-orbitals and give rise to an asymmetric exchange coupling that can be mimicked by Eq.~(\ref{ex2}). 

\begin{figure}
\centering
\centerline{\includegraphics[width=\columnwidth]{{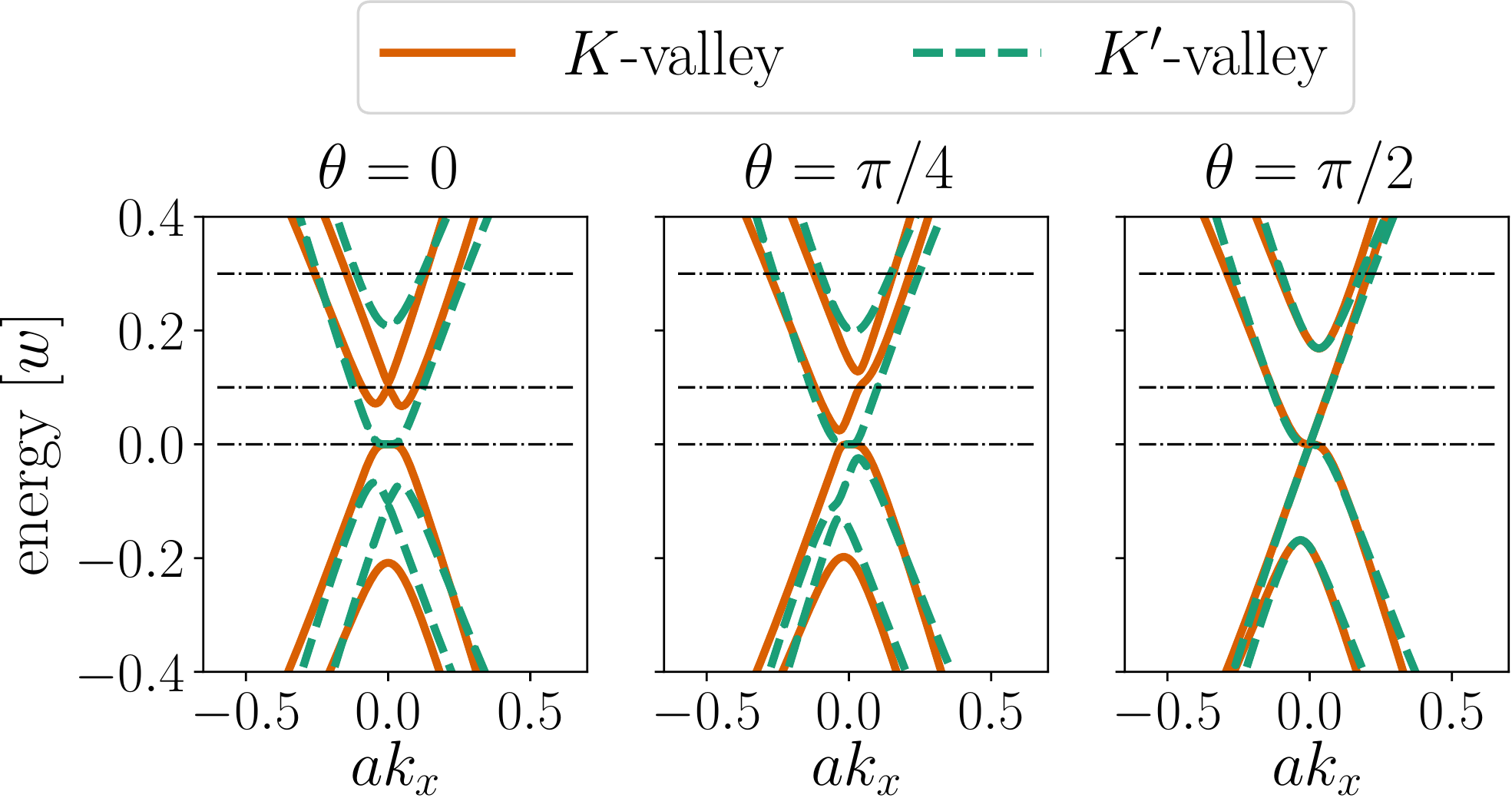}}}
\caption{Band structure for the asymmetric model. Dotted horizontal lines correspond to $E/w=0, 0.1$, and $0.3$.}
\label{fig:BSas}
\end{figure}

In order to compute non-equilibrium spin-densities we repeat the analysis of the previous sections for the coupling of Eq.~(\ref{ex2}) that we refer to below as the asymmetric model. The band-structure of the asymmetric model is illustrated in Fig.~\ref{fig:BSas}. Similarly to the symmetric model we distinguish metal and half-metal regimes that are represented now by $E=0.3\,w$ and $E=0.1\,w$, correspondingly. 

\begin{figure}
\centering
\centerline{\includegraphics[width=\columnwidth]{{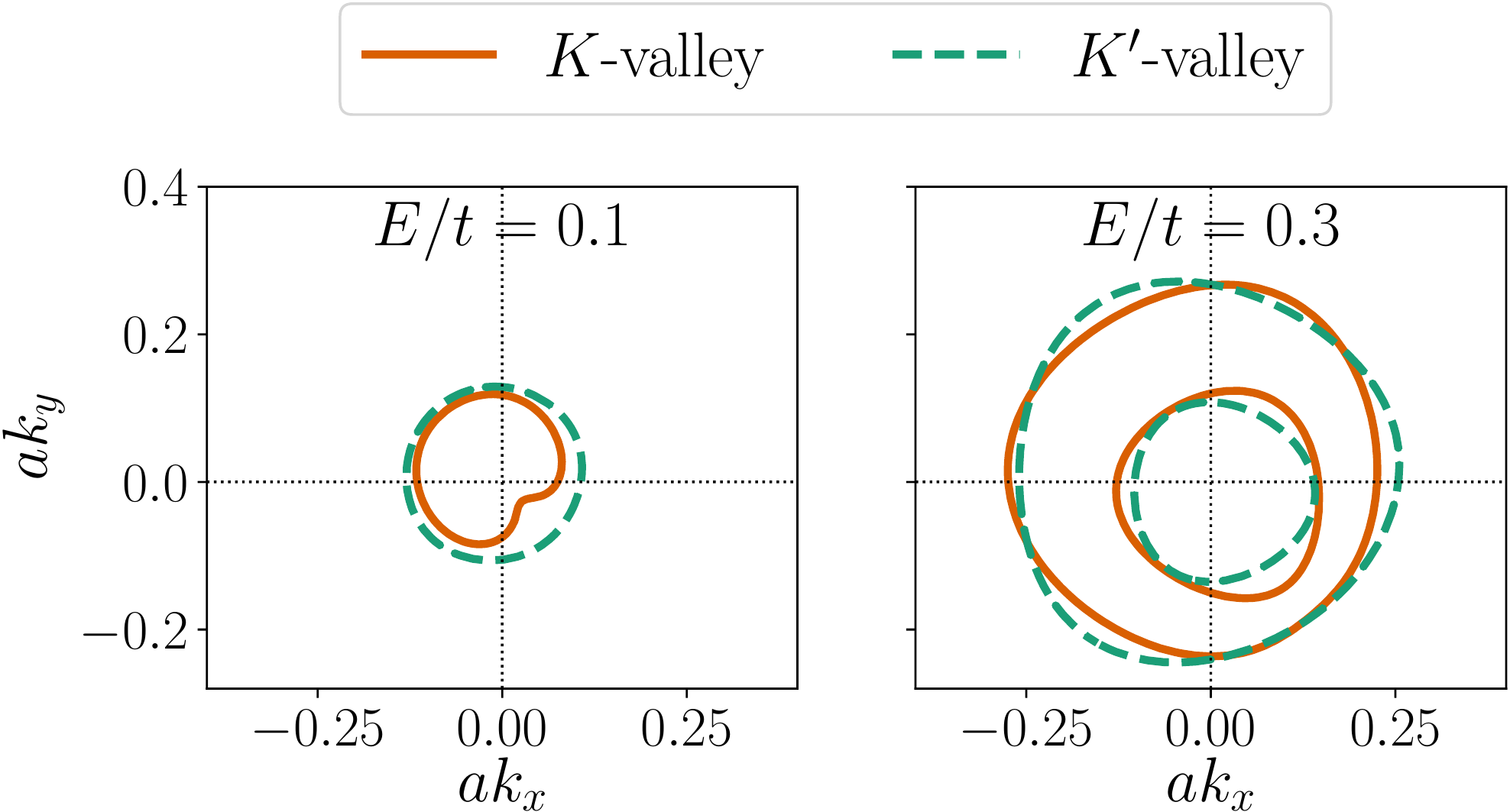}}}
\caption{The Fermi surfaces for the asymmetric model for the metal regime $E=0.3\,w$ (right panel) and the half-metal regime $E=0.1\,w$ (left panel). The Fermi surfaces are not round due to tridiagonal wrapping originating in $C_{3v}$ point-group symmetry of the crystal and due to in-plane component $\bb{n}^\textrm{A}_\parallel$ that affects the spectrum. The plots correspond to the choice $\theta=-\phi=\pi/4$. The momentum $\bb{k}$ is measured with respect to $\bb{K}$ point (solid lines) and $\bb{K}'$ point (dashed lines).}
\label{fig:FSas}
\end{figure}

The metal regime is characterized by two Fermi surfaces per valley, while the half-metal regime is characterized by a single Fermi surface per valley. The Fermi surfaces are illustrated in Fig.~\ref{fig:FSas} for $\theta=-\phi=\pi/4$, where $\theta$ and $\phi$ are now the polar and azimuthal angles of the unit vector $\bb{n}^\textrm{A}=(\cos\phi\sin\theta,\sin\phi\sin\theta,\cos\theta)^\top$, correspondingly. 

For numerical simulations we choose $\Delta= J S=0.1\,w$, $\lambda=0.05\,w$, and $V_\textrm{d}=0.5\,w$ as for the symmetric model. 

Similarly to Eq.~(\ref{low}) the low energy sector of the asymmetric model is represented by the following effective Hamiltonian
\be
\label{low2}
H^\textrm{as}_0= v\, \bb{p}\cdot\bb{\Sigma}+\alpha_\textrm{R}\lt[\bb{\sigma}\times\bb{\Sigma}\rt]_{\hat{z}} - \frac{\Delta}{2}\,\bb{n}^\textrm{A}\!\cdot\bb{\sigma}\,\lt(1+\Sigma_z\Lambda_z\rt),
\e 
that clearly lacks the sublattice symmetry. Interestingly, we observe from numerical simulations that the staggered non-equilibrium polarization $\delta \bb{s}_-$ is still vanishing in the metal regime, while it becomes finite in the half-metal regime. It has to be noted, however, that the separation to staggered and non-staggered polarization is largely irrelevant for the asymmetric model, since the latter is characterized by the only vector torque 
\be
\label{TA}
\bb{T}^\textrm{A}= \frac{J\mathcal{A}}{\hbar} \bb{n}^\textrm{A}\times\delta\bb{s}^\textrm{A}= \frac{J\mathcal{A}}{\hbar} \lt(\bb{\ell}+\bb{m}\rt)\times\lt(\delta\bb{s}_++\delta\bb{s}_-\rt),
\e
that is defined exclusively by the spin density on the sublattice A. 

Indeed, for an ``isotropic'' antiferromagnet we find, in complete analogy with Eqs.~(\ref{AFMEOM}), the equations of motion
\beml
\label{AFMEOM2}
\begin{align}
&\dot{\bb{\ell}} = -(J_\textrm{ex}/2\hbar)\, \bb{\ell}\times\bb{m} +\bb{H}\times\bb{\ell}+\bb{T}^\textrm{A},\\
&\dot{\bb{m}} = \bb{H}\times\bb{m}+\bb{T}^\textrm{A},
\end{align}
\eml
which depend only on the spin-orbit torque $\bb{T}^\textrm{A}$. 

We would like to remind that the equations of motion of Eq.~(\ref{AFMEOM2}) are essentially incomplete since we do not investigate Gilbert-damping terms and we do not take into account anisotropy of antiferromagnet (as well as Dzyaloshinskii-Moriya interaction terms). The missing terms are essential if one wants to understand the magnetization dynamics, while they cannot alter the spin-orbit torque terms that we compute. 

As one can readily see from Fig.~\ref{fig:FSas}, the Fermi surfaces in the asymmetric model are not entirely round and reveal some tri-diagonal wrapping even for energies that are close to zero, $E\ll w$. The Fermi surfaces are also shifted with respect to the points $\bb{K}$ and $\bb{K}'$ and deformed depending on the azimuthal angle $\phi$ characterizing in-plane orientation of the vector $\bb{n}^\textrm{A}$. We find, however, that despite these effects the non-equilibrium spin-density $\delta\bb{s}^\textrm{A}$ can be very well decomposed using the symmetry analysis of the group $C_{\infty v}$ in analogy to Eq.~(\ref{symmetry}) for the symmetric coupling model. 

To do that we represent the vector $\bb{n}^\textrm{A}=\bb{n}_\parallel^\textrm{A}+\bb{n}_\perp^\textrm{A}$ as the sum of in-plane and perpendicular-to-the-plane components and decompose the numerical data with respect to the transformations: $\bb{n}_\perp^\textrm{A}\to -\bb{n}_\perp^\textrm{A}$ and $\bb{n}_\parallel^\textrm{A}=-\bb{n}_\parallel^\textrm{A}$. As the result, the non-equilibrium spin density (as the function of the vector components) is readily decomposed into the sum of four contributions,  
\be
\label{dec}
\delta\bb{s}^\textrm{A}[\bb{n}^\textrm{A}_\perp,\bb{n}^\textrm{A}_\parallel]= \delta\bb{s}_{++}+ \delta\bb{s}_{+-}+\delta\bb{s}_{-+}+\delta\bb{s}_{--},
\e
which are expressed as
\begin{align}
\delta\bb{s}_{\zeta\kappa}=
&\frac{1}{4}\Big(\delta\bb{s}^\textrm{A}[\bb{n}^\textrm{A}_\perp,\bb{n}^\textrm{A}_\parallel]+\zeta\, \delta\bb{s}^\textrm{A}[-\bb{n}^\textrm{A}_\perp,\bb{n}^\textrm{A}_\parallel]\n\\
&+\kappa\, \delta\bb{s}^\textrm{A}[\bb{n}^\textrm{A}_\perp,-\bb{n}^\textrm{A}_\parallel]+\zeta\kappa\, \delta\bb{s}^\textrm{A}[-\bb{n}^\textrm{A}_\perp,-\bb{n}^\textrm{A}_\parallel]\Big), 
\end{align}
where $\zeta$ and $\kappa$ take on the values $\pm 1$. 

\begin{figure}
\centering
\centerline{\includegraphics[width=\columnwidth]{{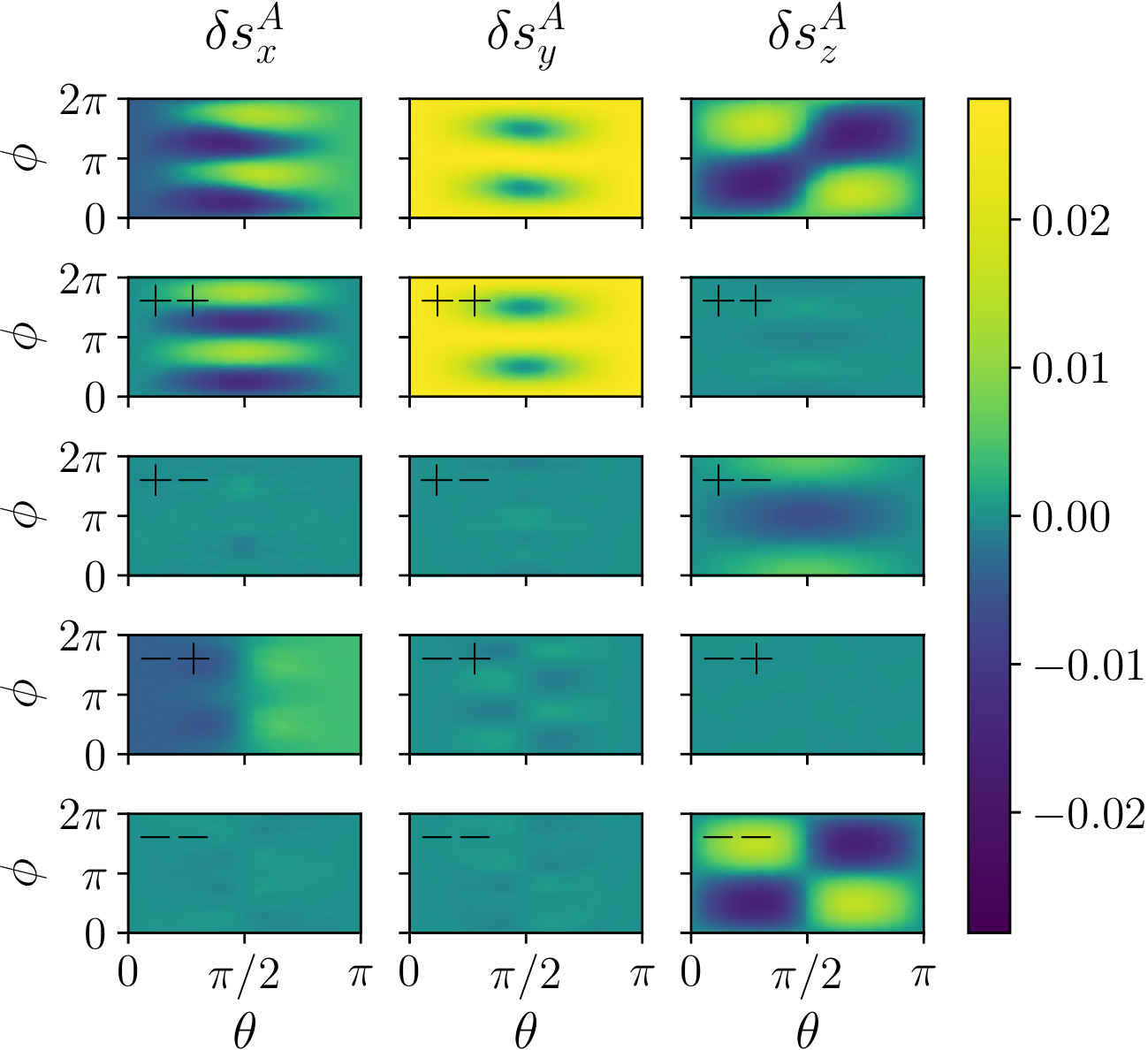}}}
\caption{Top panels show non-equilibrium spin density, $\delta\bb{s}^\textrm{A}$, for the asymmetric coupling model with subtracted background component along $\bb{n}^\textrm{A}$ in the units of $\Delta\lambda j/evw^2$. The other panels represent the results of the spin-density decomposition of Eq.~(\ref{dec}). The results presented correspond to the ansatz of Eq.~(\ref{sA}). The data for density plots have been collected for 200 different orientations of the N\'eel vector for each impurity concentration. Each data point corresponds to averaging over at least 80 impurity configurations.}
\label{fig:density_as}
\end{figure}

The bare results for non-equilibrium spin density $\delta\bb{s}^\textrm{A}$ (with a subtracted background component along $\bb{n}^\textrm{A}$) and the results for the contributions $\delta\bb{s}_{\zeta\kappa}$ are shown in Fig.~\ref{fig:density_as} for the metal regime, $E=0.3\,w$. Similar results are obtained for the half-metal regime, $E=0.1\,w$. 
 
We find that our numerical results for the non-equilibrium spin density $\delta\bb{s}^\textrm{A}$ (with subtracted background along $\bb{n}^\textrm{A}$ that does not enter the torque $\bb{T}^\textrm{A}$) are perfectly decomposed as a sum of five different vector forms in both metal and half-metal regimes,
\begin{align}
\delta\bb{s}^\textrm{A}&\,= A_I(n_z^2)\,\hat{\bb{z}}\times \bb{I} + A'_I(n_z^2)\,\bb{n}^\textrm{A}_\parallel\times \lt[\bb{n}^\textrm{A}_\parallel\times \lt[\hat{\bb{z}}\times \bb{I}\rt]\rt]\n\\
&+ B_\perp(n_z^2)\,\bb{n}^\textrm{A}_\perp\times \lt[\hat{\bb{z}}\times \bb{I}\rt]+B_\parallel(n_z^2)\,\bb{n}^\textrm{A}_\parallel\times \lt[\hat{\bb{z}}\times \bb{I}\rt]\n\\
&+C(n_z^2)\,\bb{n}^\textrm{A}_\parallel\times \lt[\bb{n}^\textrm{A}_\perp\times \lt[\hat{\bb{z}}\times \bb{I}\rt]\rt],
\label{sA}
\end{align}
where all coefficients are generally even functions of the component $n_z= n_z^\textrm{A}=\cos\theta$. Note that the first two vector forms are of the same symmetry with respect to both reflections $\bb{n}^\textrm{A}_\perp\to - \bb{n}^\textrm{A}_\perp$ and $\bb{n}^\textrm{A}_\parallel\to - \bb{n}^\textrm{A}_\parallel$.

\begin{figure}[t]
\centering
\centerline{\includegraphics[width=\columnwidth]{{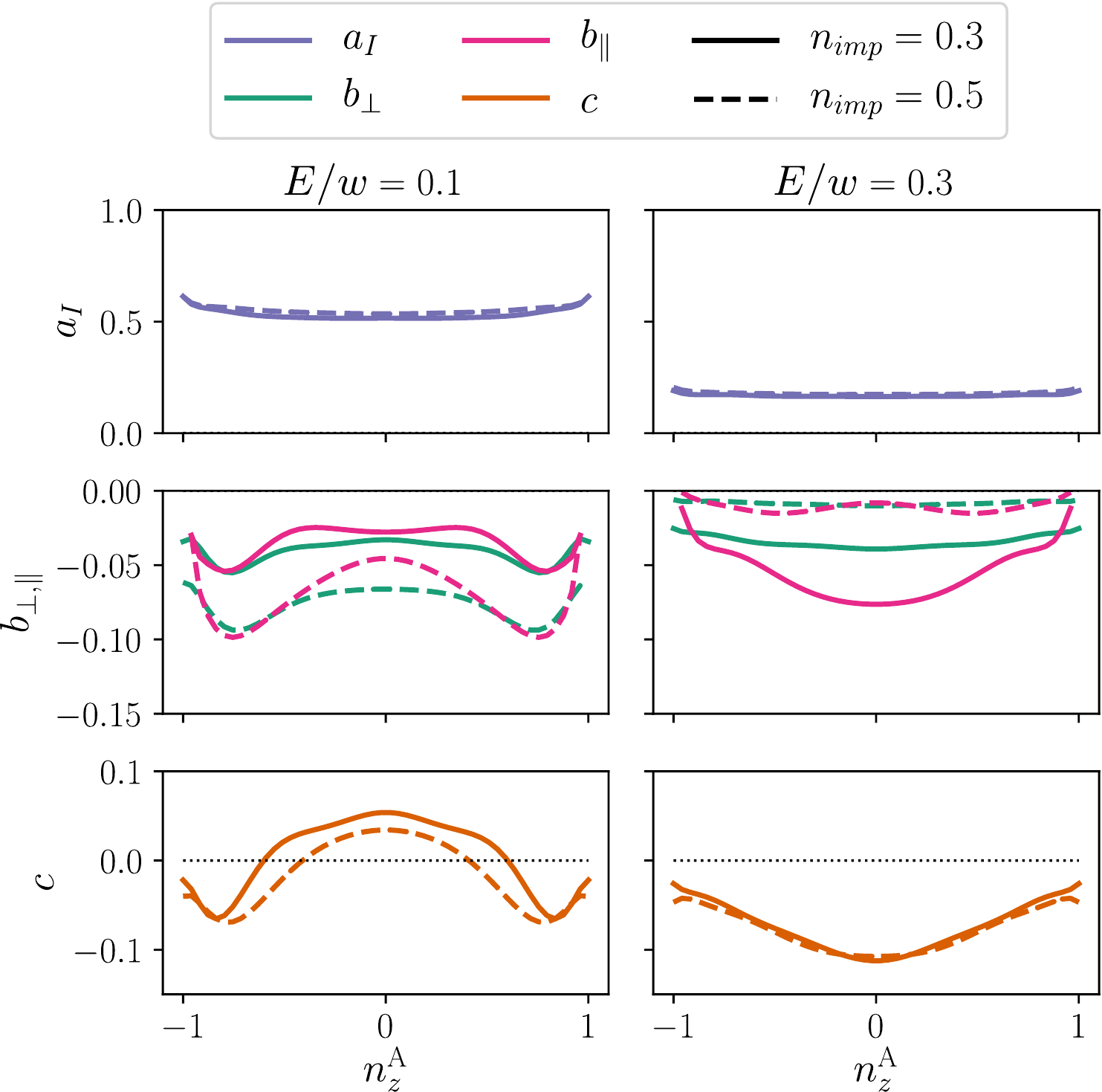}}}
\caption{Numerical results for the asymmetric coupling model that are expressed in the form of the angle dependent coefficients $a_I$, $b_\perp$, $b_\parallel$, and $c$ that define the spin-orbit torque $\bb{T}^\textrm{A}$ in Eq.~(\ref{TAresult}). The data is obtained by fitting the density data (partially represented in Fig.~\ref{fig:density_as}) using the ansatz of Eq.~(\ref{sA}).}
\label{fig:result_as}
\end{figure}

Remarkably, our numerical data again corresponds to a constant coefficient $A_I(n_z^2)=A_I$ in all regimes considered as it was also the case for the symmetric model.   

The spin orbit torque $\bb{T}^A$, which is obtained by substituting the result of Eq.~(\ref{sA}) to Eq.~(\ref{TA}), can be represented as the sum of only four vector forms 
\begin{align}
\bb{T}^\textrm{A}= \eta \Big(&a_I(n_z^2)\,\bb{n}^\textrm{A}\times\lt[\hat{\bb{z}}\times \bb{j}\rt] \n\\
&+ b_\perp(n_z^2)\,\bb{n}^\textrm{A}\times\lt[\bb{n}^\textrm{A}_\perp\times \lt[\hat{\bb{z}}\times \bb{j}\rt]\rt]\n\\
&+b_\parallel(n_z^2)\,\bb{n}^\textrm{A}\times\lt[\bb{n}^\textrm{A}_\parallel\times \lt[\hat{\bb{z}}\times \bb{j}\rt]\rt]\n\\
&+c(n_z^2)\,\bb{n}^\textrm{A}\times\lt[\bb{n}^\textrm{A}_\parallel\times \lt[\bb{n}^\textrm{A}_\perp\times \lt[\hat{\bb{z}}\times \bb{j}\rt]\rt]\rt]\Big),
\label{TAresult}
\end{align}
where we again introduce $\eta=J\mathcal{A}\Delta\lambda /evh w^2$, $\bb{j}=\bb{I}/W$ and the dimensionless coefficients $a_I$, $b_\perp$, $b_\parallel$, and $c$ that all appear to be non-trivial functions of $n_z^2=\cos^2\theta$ as shown in Fig.~\ref{fig:result_as}. Note that the function $A'_I(n_z^2)$ in Eq.~(\ref{sA}) for spin density contributes to both $a_I$ and $c$ coefficients in Eq.~(\ref{TAresult}) in the direct analogy to Eqs.~(\ref{torque_Tplus_MOD}) and (\ref{renorm}). 

Thus, the spin-orbit torque $\bb{T}^\textrm{A}$ is parameterized by $4$ dimensionless functions. These functions are shown in Fig.~\ref{fig:result_as} illustrating the main result of our numerical simulations for the asymmetric coupling model. In sharp contrast to the symmetric model we observe that the anti-damping torques are no longer vanishing. We also find that $b_\perp(n_z^2) \approx b_\parallel(n_z^2)$ in the half-metal regime. The major field-like torque $\bb{n}^\textrm{A}\times\lt[\hat{\bb{z}}\times\bb{j}\rt]$, which originates in the Edelstein effect, acquires a weak dependence on $\theta$ due to the impact of the coefficient $A'_I(n_z^2)$. 

As it is expected we observe that anti-damping torques, which are proportional to $b_{\perp,\parallel}$, depend strongly on disorder concentration, while the field-like torques do not. Moreover, while anti-damping torques are suppressed by disorder in the metal regime, the opposite is true in the half-metal regime. 

Similarly to the symmetric model, the field-like torques are smaller in the metal regime than they are in half-metal regime. One can speculate, however, that anti-damping-like torques play a leading role in the metal regime of the asymmetric coupling model for sufficiently clean samples. We also see that anti-damping torques reveal rather strong anisotropy. In the metal regime they take on maximal values for in-plane orientations of the vector $\bb{n}^\textrm{A}$. 
 
It is also worth stressing that our results for asymmetric coupling model are the same for both ferromagnetic and antiferromagnetic order, since the spin-orbit torques are insensitive to the direction of the field $\bb{n}^\textrm{B}$.  

As have been already mentioned one can expect the asymmetric model to capture also the physics of layered ferrimagnets such as GdFeCo or Pt/GdFeCo \cite{Hoffman2018, KimFERRISOT018}. The magnetization switching by means of spin-orbit torques in these materials has recently become a subject of intense studies \cite{Roschewsky2016, Roschewsky2017, KimFERRISOT018}.  In particular, the contribution from both field-like and anti-damping-like torques have been identified in GdFeCo films with perpendicular magnetocrystalline anisotropy in both transition-metal and rare-earth-metal rich configurations \cite{Roschewsky2016, Roschewsky2017}. These experimental results are, at least, qualitatively in line with our findings. 

%
\begin{figure}[t]
\centering
\centerline{\includegraphics[width=\columnwidth]{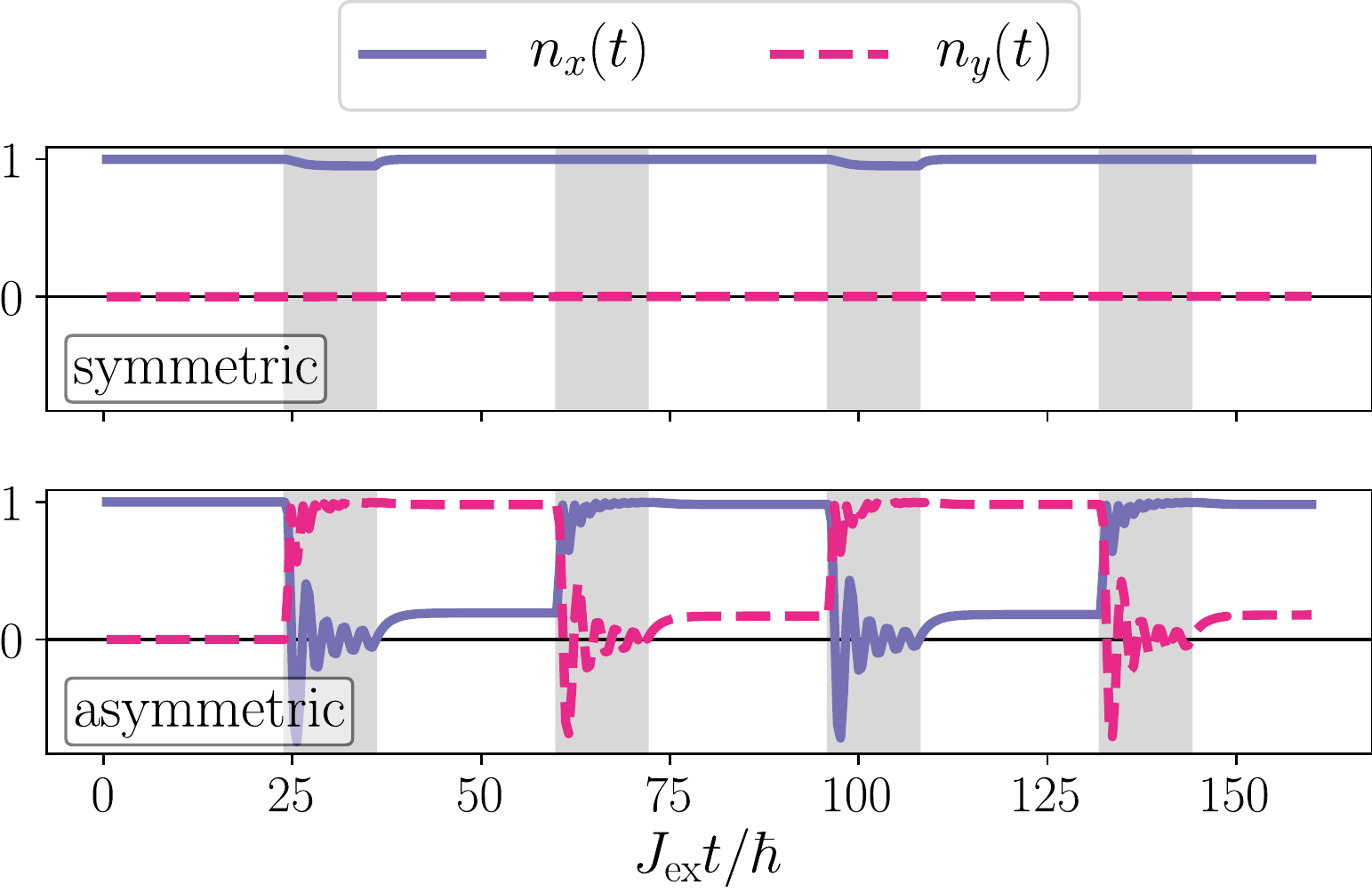}}
\caption{Time evolution of the in-plane N\'eel vector $\bb{n}_\parallel(t)$ in the symmetric model (top panel) and asymmetric model (bottom panel) induced by a sequence of current pulses alternating between $-\hat{\bb{x}}$ and $\hat{\bb{y}}$ directions (gray areas). The results are obtained from Eqs.~(\ref{EOMAFM10}) for  the parameters $\alpha=0.005$, $\hbar\eta j/J_\textrm{ex}=0.05$, $E/w=0.3$ and $n_\textrm{imp}=0.3$.}
\label{fig:switching}
\end{figure}

\section{Antiferromagnetic dynamics}

It is instructive to apply our results for the symmetric and asymmetric model to describe antiferromagnetic switching. As was mentioned in the introduction both field-like staggered torque and anti-damping-like non-staggered torque can be important for the N\'eel vector switching. In the symmetric model we only find field-like non-staggered torques that cannot induce AFM switching for realistic values of the current intensity. In the asymmetric model we find both field-like and anti-damping-like torques that act on a single sub-lattice. Although, strictly speaking, these do not entail a staggered torque, i.\,e. the N\'eel torque, one can nevertheless expect switching to occur. 

In order to model AFM magnetization dynamics we adopt the equations of motion (\ref{AFMEOM}) for the unit vectors $\bb{n}^\textrm{A}$ and $\bb{n}^\textrm{B}$  with Gilbert damping terms included,
\beml
\label{EOMAFM10}
\begin{align}
\dot{\bb{n}}^\textrm{A} &= \frac{J_\textrm{ex}}{2\hbar}\bb{n}^\textrm{B}\times \bb{n}^\textrm{A}+ \frac{J\mathcal{A}}{\hbar}\, \bb{n}^\textrm{A}\times \bb{s}^\textrm{A} +\alpha\, \bb{n}^\textrm{A}\times \dot{\bb{n}}^\textrm{A},\\
\dot{\bb{n}}^\textrm{B}&= \frac{J_\textrm{ex}}{2\hbar} \bb{n}^\textrm{A}\times \bb{n}^\textrm{B}+ \frac{J\mathcal{A}}{\hbar}\, \bb{n}^\textrm{B}\times \bb{s}^\textrm{B}+\alpha\, \bb{n}^\textrm{B}\times \dot{\bb{n}}^\textrm{B},
\end{align}
\eml
where we take $\alpha=0.005$ for the strength of the Gilbert damping.  

In Fig.~\ref{fig:switching} we illustrate the AFM dynamics induced by short current pulses of the density $j=0.05\,J_\textrm{ex}/\eta\hbar$, which are applied in alternating directions $-\hat{\bb{x}}$ and $\hat{\bb{y}}$. The parameters of spin-orbit torques are derived for the Fermi energy $E =0.3\,w$ (in the metal regime) and for the impurity concentration $n_\textrm{imp}=0.3$ which corresponds to placing impurities on $30\%$ of lattice sites.   

The spin-orbit torques in the symmetric model are obtained by substituting $\bb{s}^\textrm{A}=\bb{s}^\textrm{B} = \delta\bb{s}_+$ in Eqs.~(\ref{EOMAFM10}), where $\delta\bb{s}_+ \propto \hat{\bb{z}}\times\bb{j}$ is given by Eq.~(\ref{s_SOT}). The spin-orbit torques in the asymmetric model correspond to $\bb{s}^\textrm{B}=0$ and $\bb{s}^\textrm{A}$ taken from Eq.~(\ref{sA}).
 
It is worth noting that the switching behaviour in the asymmetric model (shown in the bottom panel of Fig.~\ref{fig:switching}) is provided exclusively by the field-like-torques (which are proportional to the $a_I$ and $c$ in Eq.~(\ref{TAresult})). In contrast, anti-damping torques, which are proportional to $b_{\perp,\parallel}$, do not play an important role in this behaviour for the proposed choice of parameters. Nevertheless, if anti-damping torques become essentially enhanced to overcome Gilbert damping they may lead to qualitatively different behaviour. Indeed, for giant anti-damping torques, the spin angular momentum in the asymmetric model is transferred from one sub-lattice to the another. Such a process, however, destroys the N\'eel order and fully magnetizes the sample during the current pulse. Once the electric current is switched off the N\'eel order is restored albeit in a different direction. This type of switching bears some resemblance to all-optical ultra-fast magnetization switching that is observed, for example, in a ferrimagnet GdFeCo \cite{RevModPhys.82.2731}. In the latter case the crystal also enters a transient fully magnetized state during a femtosecond optical pulse and relaxes to a state with opposite magnetization soon after the pulse.  We note however, that a current-induced switching of this type in our AFM model would require unrealistically large charge current intensities due to rather small magnitude of anti-damping torques. 

\section{Conclusions}

Motivated by recent experiments on the N\'eel vector switching we investigate microscopically the spin-orbit torques in an $s$-$d$-like model of a two-dimensional honeycomb antiferromagnet with Rashba spin-orbit coupling. We investigated the model with preserved and broken sublattice symmetry and distinguished metal and half-metal regimes for each of the model. Spin-orbit interaction in combination with on-site disorder potential and local exchange coupling between conduction and localized spins have been responsible for a microscopic mechanism of the angular momentum relaxation. We find identically vanishing anti-damping and N\'eel spin-orbit torques in the symmetric model in all regimes considered. As the result, the metal regime of the symmetric model is characterized by a particularly simple isotropic field-like spin-orbit torque, while the half-metal regime is characterized by anisotropic spin-orbit torques of the field-like symmetry.  Finite and anisotropic anti-damping torques, which are essentially dissipative and, therefore, depend on disorder strength, are found in both metal and half-metal regimes of the asymmetric model. We also find non-equilibrium staggered polarization in the half-metal regime of the asymmetric model. This formally leads to a finite value of the N\'eel spin-orbit torque, that is, however, not a useful quantity in the asymmetric model.  Overall, our results reveal the importance of two-dimensional electron momentum confinement for spin-orbit torque anisotropy. Largest values of spin-orbit torques are also associated with the half-metal regimes of conduction in both models.

\begin{acknowledgments}
This research was supported by the JTC-FLAGERA Project GRANSPORT.  D.Y. and M.T. acknowledge the support from the Russian Science Foundation Project No. 17-12-01359. The work of DY was also supported by the Swedish Research Council (Vetenskapsr{\aa}det, 2018-04383). A.M. and S.G. were supported by the King Abdullah University of Science and Technology (KAUST).
\end{acknowledgments}

\bibliographystyle{apsrev4-1}
%

\end{document}